\documentclass[twocolumn]{aastex62}
\usepackage{epsfig}
\usepackage{graphicx}
\usepackage{float}
\usepackage{amsmath}
\usepackage{color}
\usepackage{amssymb}
\usepackage{amsfonts}
\usepackage{units}
\usepackage{amsmath,bm}
\usepackage{amssymb}
\usepackage{amsfonts}
\usepackage{subfigure}
\usepackage{etoolbox}
\usepackage{appendix}
\usepackage{booktabs}
\usepackage{lineno}

\newcommand{\be}{\begin{equation}}
	\newcommand{\ee}{\end{equation}}
\newcommand{\ba}{\begin{eqnarray}}
	\newcommand{\ea}{\end{eqnarray}}
\newcommand{\om}{\omega}

\begin{document}
%	\linenumbers
	
	\title{Polarization Evolution of Fast Radio Burst Sources in Binary Systems}
	
	\author{Zhao-Yang Xia}
	\affil{School of Astronomy and Space Science, Nanjing University, Nanjing 210023, China}
	\affil{Key Laboratory of Modern Astronomy and Astrophysics (Nanjing University), Ministry of Education, Nanjing 210023, China}
	
	\author{Yuan-Pei Yang}
	\affil{South-Western Institute for Astronomy Research, Yunnan University, Kunming 650500, China; ypyang@ynu.edu.cn}
	\affil{Purple Mountain Observatory, Chinese Academy of Sciences, Nanjing, Jiangsu 210023, China}
	
	\author{Qiao-Chu Li}
	\affil{School of Astronomy and Space Science, Nanjing University, Nanjing 210023, China}
	\affil{Key Laboratory of Modern Astronomy and Astrophysics (Nanjing University), Ministry of Education, Nanjing 210023, China}

	\author{Fa-Yin Wang}
	\affil{School of Astronomy and Space Science, Nanjing University, Nanjing 210023, China}
	\affil{Key Laboratory of Modern Astronomy and Astrophysics (Nanjing University), Ministry of Education, Nanjing 210023, China}

	\author{Bo-Yang Liu}
	\affil{South-Western Institute for Astronomy Research, Yunnan University, Kunming 650500, China}

	\author{Zi-Gao Dai}
	\affil{Department of Astronomy, School of Physical Sciences, University of Science and Technology of China, Hefei 230026, China; daizg@ustc.edu.cn}
	%\affil{School of Astronomy and Space Science, Nanjing University, Nanjing 210023, China}

	\begin{abstract}
		Recently, some fast radio bursts (FRBs) have been reported to exhibit complex and diverse variations in Faraday rotation measurements (RM) and polarization, suggesting that dynamically evolving magnetization environments may surround them.  In this paper, we investigate the Faraday conversion (FC) effect in a binary system involving an FRB source and analyze the polarization evolution of FRBs. For an strongly magnetized high-mass companion binary (HMCB), when an FRB with $\sim100\%$ linear polarization passes through the radial magnetic field of the companion star, the circular polarization (CP) component will be induced and oscillate symmetrically around the point with the CP degree equal to zero, the rate and amplitude of the oscillation decrease as the frequency increases. The very strong plasma column density in the HMCBs can cause CP to oscillate with frequency at a very drastic rate, which may lead to depolarization.
		Near the superior conjunction of the binary orbit, the DM varies significantly due to the dense plasma near the companion, and the significant FC also occurs in this region.
		As the pulsar moves away from the superior conjunction, the CP gradually tends towards zero and then returns to its value before incidence.
		We also investigate the effect of the rotation of the companion star. We find that a sufficiently significant RM reversal can be produced at large magnetic inclinations and the RM variation is very diverse. Finally, we apply this model to explain some polarization observations of PSR B1744-24A and FRB 20201124A.
	\end{abstract}
	
	\section{Introduction}
	Fast radio bursts (FRBs) are mysterious intense radio transients with millisecond duration at cosmological distances   \citep[for reviews, see][]{Petroff19, Cordes19,zha20b, Xiao21,zhang20,zhang22}.
	To date, a few hundred FRB sources have been detected \citep[e.g.,][]{lor07, tho13, spi16, cha17,ban19}. Most of them were found to be one-off events, and over 50 FRB sources showed repeating activities\citep{Chime23}.
	However, the origin of FRBs is still a big puzzle.
	Remarkably, FRB 20200428 was detected to be associated with an X-ray burst from the Galactic magnetar SGR 1935+2154   \citep{boc20,chi20b, mer20,li21,tav21}, which established an FRB–magnetar connection.
	
	The evolutions of dispersion measurement (DM), Faraday rotation measurement (RM), and polarization are important clues to constrain the physical origins of FRBs and their surrounding environment.
	The first observed repeating FRB 20121102A  \citep{Mich18}, has an RM of up to $10^5$ $ \rm rad~m^{-2}$, which is the largest of all FRBs so far.
	The RM of this source decreases $\sim 15\%~\unit{yr^{-1}}$   \citep{hilmarsson21}, while the DM increases with $d\text{DM}/dt \simeq 0.85 \,\unit{pc\,cm^{-3}\,yr^{-1}}$ over a six year  \citep{lidi21, hilmarsson21}. Most bursts from this source showed strong linear polarization (LP). However, the observation of Five-hundred-
	meter Aperture Spherical radio Telescope (FAST) recently revealed that a dozen bursts in nearly 2000 bursts behavior significant circular polarization(CP) with the highest one reaching the CP degree of 64$ \% $ \citep{Feng22}. Besides, it is interesting that both FRB 20121102A and FRB 20190520B appeared significant frequency-dependent depolarization, which might originate from the multi-path propagation \citep{Feng22b, Yang22b}.

	Another repeater, FRB 20190520B has a very large host DM of about 900 $\rm pc~cm^{-3}$  \citep{niu22, Ocker22}, which is nearly an order of magnitude higher than the other FRBs. The source also has an RM of up to $ 10^4 ~\rm rad~m^{-2}$, which is the second largest RM among all observed FRB sources, and showed an RM sign reversal over six months   \citep{Anna23}, accompanied by fluctuations in the CP over a range of  $\pm 10 \% $   \citep{Feng22}, while the DM of the source has remained almost constant over this period.
	
	The extremely active repeater,  FRB 20201124A  also showed significant and irregular short timescale RM variations during the first 36 days of the observation period, then the RM value suddenly became stable during the next 18 days   \citep{xu22}. This source also has a significant CP up to 75 $\%$ \citep{xu22, Jiang22}, and the CP and LP are quasi-periodically oscillating with frequency in a small fraction of the bursts (e.g., the bursts 779 and 926). This oscillation disappears when the RM variation stops (e.g., burst 1472).
	% On the other hand, a good proportion, close to 50 $ \% $  \citep{Feng22}, of the non-repetitive FRBs detect  CP  \citep{masui15, petroff15, caleb18}, which is much larger than the fraction of pulses having CP in the repeating FRBs.
	Thus, the significant and diverse variations in RMs polarisations of some FRB repeaters imply the existence of a dynamically evolving magnetized environment around the FRB source \citep{wang22}.
	
	There is some evidence suggesting that some FRBs might be in binary systems.
	First, the repeating FRB 20180916B has 16.35 days of periodic activity \citep{Chime2020b}, and FRB 20121102A has a possible 160-day period \citep{Cruces21, Rajwade20}, which may correspond to the orbital period of a binary star system.
	Second, there is $\sim 250\,\unit{pc}$ offset between FRB  20180916B and the brightest region of the nearest young stellar clump in its host galaxy, the possible birthplace of FRB 20180916B  \citep{Tendulkar21}. This indicates that the age of the central object is more than $10^5\,\unit{yr}$. It is inconsistent with the scenario involving a young magnetar  \citep{ten21} but seems to agree with scenarios of the high-mass X-ray binaries  \citep{bod12}.
	Third, FRB 20180916B and FRB 20201124A show significant RM variations \citep{mckinven22, xu22}, which can be explained by the magnetic field reversal of the binary star along the line of sight  \citep{wang22,zhao23}.
	Fourth, FRB 20200120E was found to be located in a globular cluster \citep{Kirsten22}, which is known to be rich in binaries from many dynamical interactions.
	Fifth,
	FRB 20201124A has been detected to appear some changes in CP, which might be explained by FC  \citep{xu22} near a highly magnetized companion. This is very similar to PSR B1744-24A in a binary system, meanwhile, it also has been observed to appear irregular large RM variation  \citep{li23}.
	
\begin{figure*}
		\centering
		\subfigure[Magnetized Companion with a dipolar field.]{
			\includegraphics[width=0.4\linewidth]{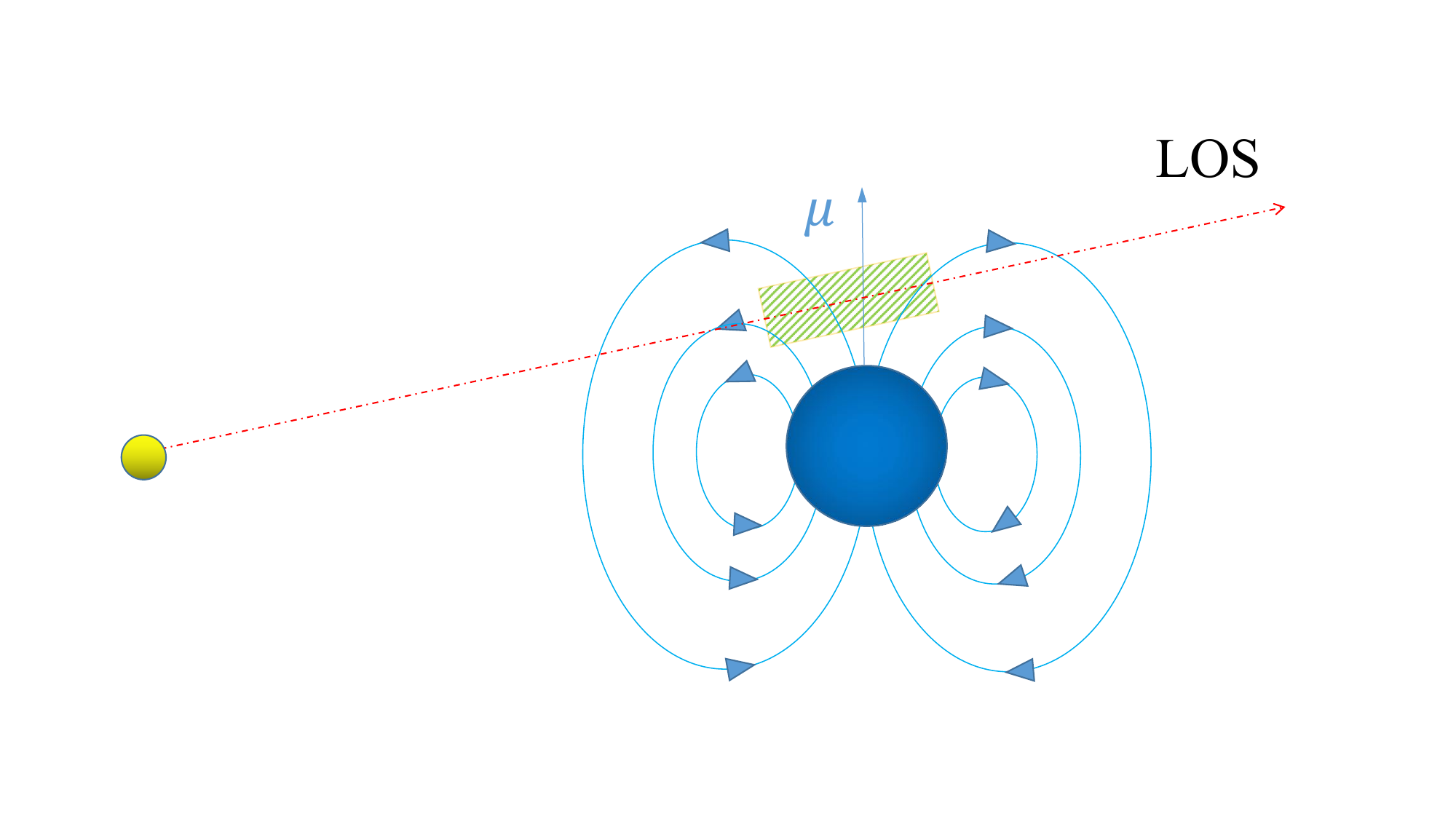}
			\label{fig dipolar}
		}
		\subfigure[Mangetized Companion with a radial field.]{
			\includegraphics[width=0.4\linewidth]{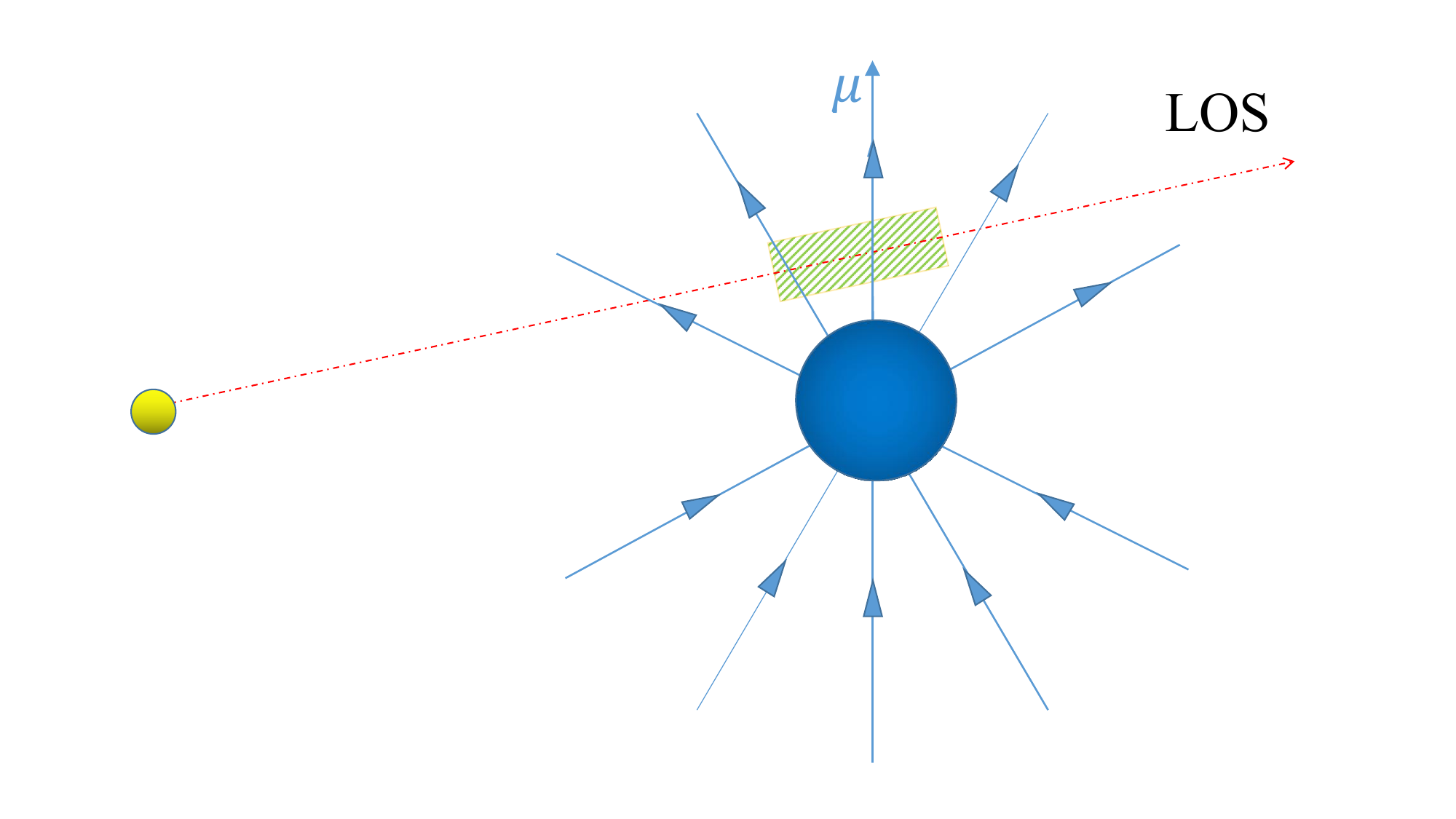}
			\label{fig radial}
		}
		\caption{The side view of a binary system involving neutron stars with different companion magnetic field configurations, where the yellow circular is the neutron star, which is at the superior conjunction (the neutron star behind the companion star), and the rQT region is likely to be present in both scenarios, as shown in the shaded area of the figure.}\label{fig rQT}
	\end{figure*}

	In this paper, we investigate in detail the FC effect in cold plasmas and apply it to the binary system involving an FRB source, and analyze the polarization evolution of the radio bursts from the FRB source.
	This paper is organized as follows.
	In Section \ref{sec:descri}, we describe the propagation behavior of electromagnetic (EM)  waves in the local magnetized cold plasma and discuss the FC effect in detail.
	In section \ref{sec:Binary}, we calculate the radio wave propagation in the rQT region in these two scenarios illustrated in Fig. \ref{fig rQT}, and based on this, some relevant observations in FRBs and radio pulsars are explained.
	Finally, in Section \ref{sec:summary}, we summarise our conclusions and discuss their implications.

	%--------------------------------------------------------
	\section{Description of polarisation wave radiative transfer} \label{sec:descri}
	%--------------------------------------------------------
	
	First, we describe the polarization evolution of EM waves as they propagate through the local magnetized cold plasma.
	Physically, the EM wave could be decomposed into two natural modes \footnote{See Appendix \ref{cold plasma equations} for details on the derivation of the dispersion relations of the two natural modes, and the expressions correspond to the positive and negative signs in  Equation (\ref{pm n delta}).} when it propagates in a local magnetized plasma region, and we will discuss the polarization properties of these two natural wave modes in this section.

	\subsection{Polarization of the natural wave modes}\label{sec:polarization}
	In EM theory, it is customary to study polarization in the coordinate system of $ \boldsymbol{k}  $ along the $z$-axis.
	With  $ \boldsymbol{k}=\hat{\boldsymbol{x}} k_{x}+\hat{\boldsymbol{z}} k_{z} $, say that  $ E_{x^{\prime}}=\boldsymbol{E} \cdot(\hat{\boldsymbol{y}} \times \hat{\boldsymbol{k}}) $  and $  E_{y^{\prime}}=\boldsymbol{E} \cdot \hat{\boldsymbol{y}}  $, where  $ \hat{\boldsymbol{k}}=\boldsymbol{k} / k  $   is the unit wave vector. This introduces an alternative coordinate system $ x^{\prime}y^{\prime}z^{\prime}  $,  in which  $ \boldsymbol{k} $ is along the  $ z^{\prime} $ axis while $ \boldsymbol{B} $ is in the $ x^{\prime}z^{\prime} $ plane, yielding  $ \hat{\boldsymbol{k}} \times \hat{\boldsymbol{B}}=-\sin \theta    \, \hat{\boldsymbol{y}}  $.
	It is worth emphasizing that even though the polarization is dependent on the chosen coordinate system, the dispersion relation remains Lorentz invariant in any coordinate system.

	The coordinate transformations of the tensor satisfy $ \stackrel{\leftrightarrow}{\boldsymbol{\varepsilon}^{\prime}}=\boldsymbol{Q} \cdot 	\stackrel{\leftrightarrow}{\boldsymbol{\varepsilon}} \cdot \boldsymbol{Q}^{T}  $, where $ \boldsymbol{Q} $ is the rotation matrix, which in this context  takes the form
	\begin{equation}
		\boldsymbol{Q}=\left[\begin{array}{rcc}
			\cos \theta &  0 &  -\sin \theta
			\\
			0 & 1 & 0
			\\
			\sin \theta&  0 & \cos \theta
		\end{array}\right].
	\end{equation}	
	
	Then we can write the wave equation in the $x^{\prime}y^{\prime}z^{\prime}$ coordinate system
	\begin{eqnarray}
		\left[\begin{array}{ccc}
			S \cos ^{2} \theta+P\sin ^{2} \theta -n^2 &~~ -i D \cos \theta & ~~\left(S-P\right) \sin \theta \cos \theta \\
			i D \cos \theta&~~ S-n^2 & ~~i D \sin \theta \\
			\left(S-P\right) \sin \theta\cos \theta &~ -iD \sin \theta & ~S \sin ^{2} \theta+P \cos ^{2} \theta
		\end{array}\right] \nonumber\\
		\times\left[\begin{array}{c}
			E_{x^{\prime}} \\
			E_{y^{\prime}} \\
			E_{z^{\prime}}
		\end{array}\right]=0 \quad. \label{ExEyEz}
	\end{eqnarray}

	The electric vector can be divided into transverse and longitudinal components. The longitudinal part corresponds to an electrostatic oscillating wave. We are mainly interested in weakly anisotropic medium  \citep{Melrose91}, which refers to the fact that EM waves could be split into two natural wave modes ($ \mathit{anisotropic} $) when they propagate in the medium, and the difference in ray paths between these two components is not significant ($ \mathit{weak} $, and $n^2\simeq 1$). Thus, we can ignore the effect of wave refraction.
	However, the refractive index difference between the two natural wave modes is still important, because it leads to the evolution of the relative phase between the two components.
	When these two components are recombined, the polarization state of the EM wave usually differs compared to that before incidence.  Under the weak anisotropy approximation, the longitudinal part of the polarization is negligible  \citep{Melrose04}.
	Therefore, we only require to be concerned with the transverse part of the natural wave mode polarization, whose polarization ellipse can be completely described by the axial ratio
	\begin{eqnarray}
		T=\frac{i E_{x^{\prime}}}{E_{y^{\prime}}}=\frac{D P\cos \theta}{An^2-PS}, \label{iEx/Ey}
	\end{eqnarray}
	where $ T=\pm 1 $ corresponds to the opposite CP while $ T=0,\infty $  corresponds to the mutually orthogonal LP. It is obtained by combining the first and third lines of  Equation (\ref{ExEyEz}), which provides the correlation between the axial ratio $T$ of the polarization ellipse and the refractive index $n$, consequently substituting it into the dispersion relation (\ref{n^2}) will give
	\begin{equation}
		T^{2}+2 \mathcal{R} T-1=0 \label{T^2},
	\end{equation}	
	where $ \mathcal{R} $ is defined as the polarization parameter and its detail expression is
	\begin{equation}
		\mathcal{R}=\frac{Y\left(1-Y^2-X+\epsilon^2 X\right) \sin ^{2} \theta}{2\epsilon(1-X)(1-Y^2) \cos \theta},  \label{R/2}
	\end{equation}
	where   $X=\omega_{\rm p}^2/\omega^2$ and $Y=\omega_{\rm B}/\omega$ are two dimensionless parameters  incorporating $\omega_{\rm p}$ and $\omega_{\rm B}$.
	It is very striking that it is the same form as the one required by the two approximations\footnote{By comparing the values of the two terms in the numerator of the dispersion relation, two approximate conditions can be given, which can simplify the dispersion relation. A detailed discussion is in Appendix \ref{cold plasma equations}.} we defined in  Equations (\ref{pm QT}) and (\ref{pm QL}). This infers that $ |\mathcal{R}|\gg 1 $ corresponds to the quasi-transverse  (QT) approximation while $  |\mathcal{R}| \ll 1 $  corresponds to the quasi-longitudinal  (QL) approximation. Next, we will further investigate the connection between the polarization parameters and the ellipticity of the natural modes.

	The explicit  solution of the quadratic equation  (\ref{T^2}) satisfied by the axis ratio $ T $ is
	\begin{equation}
		T=T_{\pm}=  -\mathcal{R}\pm\left( \mathcal{R}^2+1 \right)^{1/2}, \label{T pm}
	\end{equation}
	leading to $	T_{+} T_{-}=-1 $, which implies that the polarization ellipses of the two natural wave modes are orthogonal to each other.
	The LP degree is defined as $\Pi_{l}=(T_{\pm}^2-1)/(T_{\pm}^2+1)$ and the CP degree as $\Pi_{c}=2T_{\pm}/(T_{\pm}^2+1)$. More generally, the wave mode ellipticity angle $\chi_{\rm B}  $ can be given by the axial ratio, that is $ 	\tan \chi_{\rm B}= -T_{-} $
	and $ 	\cot \chi_{\rm B}= T_{+} $ or expressed directly in terms of the polarization parameter with $ 	\tan \chi_{\rm B} =\mathcal{R}+ {\rm sgn}(\mathcal{R})\sqrt{ \mathcal{R}^2+1 }$  \citep{Broderick10, Melrose91}.

	For the QT approximation (\ref{pm QT}), which is equivalent to the one for $ \mathcal{R}\gg 1 $, we can immediately determine that $ T_{\mathrm{O}}  \simeq \infty $, $ T_{\mathrm{X}} \simeq  0 $, which means that the two natural wave modes are completely linearly polarized along the $ x^{\prime} $ and $ y^{\prime}  $ axes, respectively. The subscripts are taken according to the customary definition that the polarization of the O-mode is in the plane of  $ \boldsymbol{B} $ and  $ \boldsymbol{k} $, while the polarization of the X-mode is perpendicular to this plane.
	It is worth noting that for a pure pair plasma, $\epsilon=0$, all propagation directions satisfy the QT approximation, which means that in this case, the two natural modes of the plasma are always orthogonally linearly polarized.  Electrons and positrons contribute the same sign to the LP component but the opposite sign to the CP component,  leading to CP originating only from the asymmetric distribution of electrons and positrons in the pair plasma.
	On the other hand, for the QL approximation (\ref{pm QL}),  $ \mathcal{R}\ll 1 $, one has $ T_{\mathrm{R}}  \simeq +1 $ while the other has $ 	T_{\mathrm{L}} \simeq  -1 $, which corresponds to two CP with opposite handedness.

	%--------------------------------------------------------
	\subsection{Faraday Conversion}
	The evolution of the polarization of an EM wave is determined entirely by its two natural wave modes decomposed in the medium,  which can be described by the transfer equation for the Stokes parameters  \citep{Melrose91}
	\begin{eqnarray}
		\frac{d}{d z}\left(\begin{array}{l}
			Q \\
			U \\
			V
		\end{array}\right)=\left(\begin{array}{ccc}
			0 & -\rho_{\rm V} & \rho_{\rm U} \\
			\rho_{\rm V} & 0 & -\rho_{\rm Q} \\
			-\rho_{\rm U} & \rho_{\rm Q} & 0
		\end{array}\right)\left(\begin{array}{l}
			Q \\
			U \\
			V
		\end{array}\right), \label{transfer equation array0}
	\end{eqnarray}
	here we assume that the polarized wave\footnote{Note that the EM wave described by  Equation (\ref{transfer equation array0}) is fully polarized.} propagates along the $ z $-direction.
	An alternative way to write  Equation (\ref{transfer equation array0}) is
	\begin{equation}
		\frac{d \boldsymbol{P}}{d z}=\boldsymbol{\rho} \times \boldsymbol{P},  \label{rho ro P}
	\end{equation}
	where  $\boldsymbol{\rho}= (\rho_{\rm Q},\rho_{\rm U},\rho_{\rm V} ) $ is  the eigenvectors of the square matrix in  Equation (\ref{transfer equation array0}), which corresponds to  one of the
	two natural modes,  and $ ( Q, U, V) $ describes the polarization state  of the EM wave.

\begin{figure}
		\centering
		\includegraphics[width=0.8\linewidth]{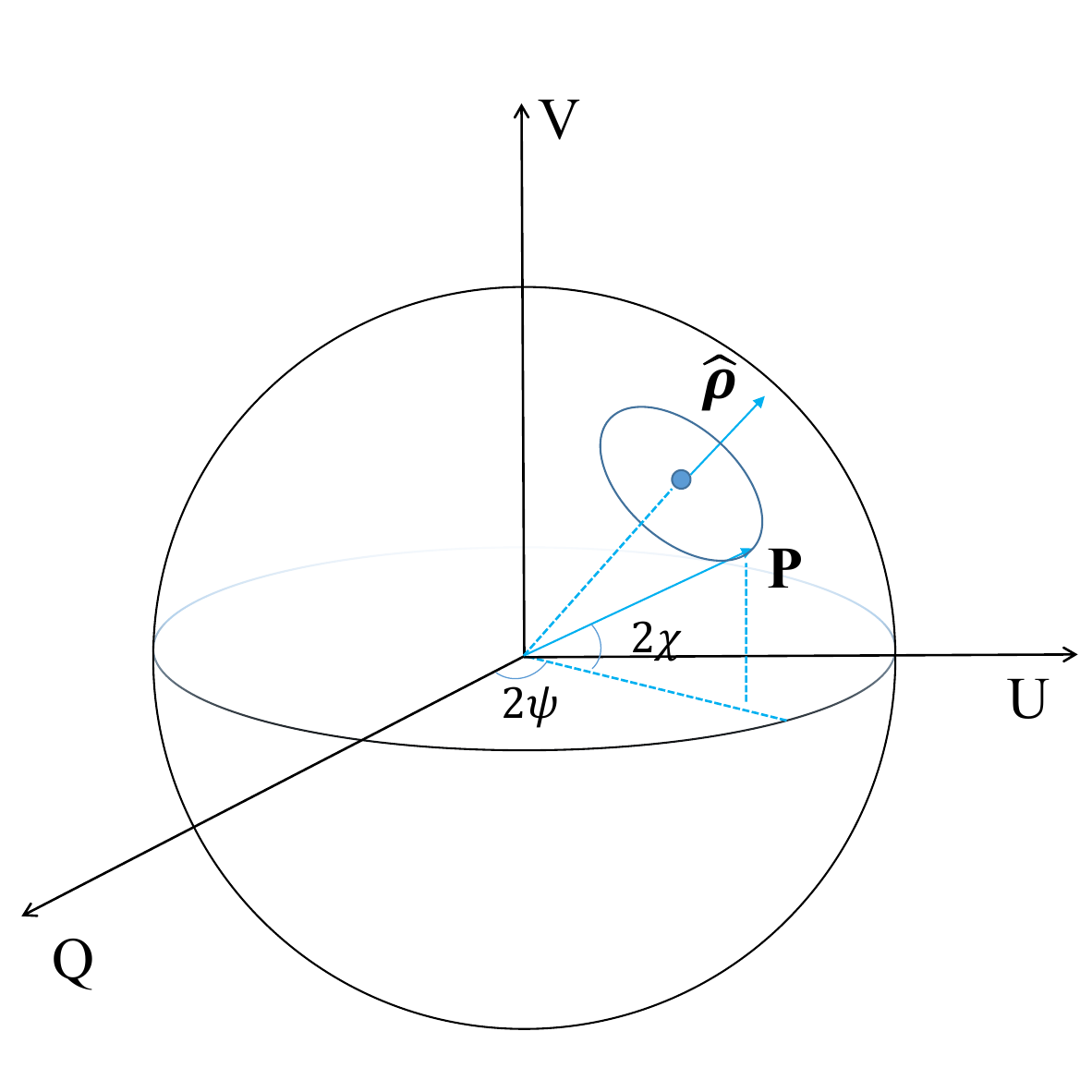}
		\caption{A geometric interpretation of the generalized Faraday rotation by the Poincaré sphere. The polarization state  $ \boldsymbol{P}=(Q, U, V) $ can be determined  by  latitude $ 2\chi $ and longitude $ 2\psi $. The generalized Faraday rotation corresponds to  the polarization point $\boldsymbol P $ rotating around the modal axis $ \boldsymbol \rho $ at constant latitude.}
		\label{Poincare}
	\end{figure}

	To better understand the physical meaning of  Equation (\ref{transfer equation array0}), one usually introduces the concept of the Poincare sphere (Fig. \ref{Poincare}). The Poincaré sphere represents a polarization state as a unit vector  $ \boldsymbol{P} $ determined by both latitude $ 2\chi $ and longitude $ 2\psi $,  which characterize the ellipticity of the polarized wave and the relative phase of its LP component, respectively. The north ($ V=1 $) and south ($ V=-1 $) poles of the sphere represent the right and left CP, respectively, while the equator ($ \chi=0$)  corresponds to complete LP.
	The polarization of the natural wave modes are mutually orthogonal, as shown by  Equation (\ref{T pm}). They can be represented by two diametrically opposite points ($\pm 2\chi_{\rm B}, \pm 2\psi_{\rm B}$ ) at two ends of the diameter axis through the center of the Poincaré sphere. This axis is called the modal axis.
	Together with  Equation (\ref{rho ro P}), the Poincaré sphere provides a geometric interpretation of the generalized Faraday rotation. It shows that the polarization point $ P $ rotates around the modal axis at a constant latitude, and the speed of rotation  $\rho$   is determined by the natural mode properties of the medium.

	The parameters $ \rho_{\rm Q}$, $ \rho_{\rm U}$ and $\rho_{\rm V} $ are calculated from the dielectric tensor of the plasma (which is exactly equivalent to the natural wave modes we derived on the previous section),  and their detailed expressions are
	\begin{eqnarray}
		\rho_{\rm Q}&=&-\Delta k \cos 2 \chi_{\rm B} \cos 2 \psi_{\rm B},\nonumber \\
		\rho_{\rm U}&=&-\Delta k \cos 2 \chi_{\rm B} \sin 2 \psi_{\rm B}, \nonumber \\
		\rho_{\rm V}&=&-\Delta k \sin 2 \chi_{\rm B},
	\end{eqnarray}
	where $|\boldsymbol \rho|= \Delta k $ is the difference in wavenumber between the two natural modes, while $ 2 \chi_{\rm B} $ is the latitude and $ 2 \psi_{\rm B} $ is the longitude of the natural mode on the Poincaré sphere, and  $  \chi_{\rm B} $, $  \psi_{\rm B} $ correspond to the ellipticity of the natural mode and the relative phase of the ray to the magnetic field, respectively.

	For the ion-electron plasma (described by $\epsilon=1$) under the weak anisotropy approximation, following  Equation (\ref{n^2}) we can derive the specific form of the wavenumber difference between the two natural modes
	\begin{equation}
		\Delta k=\frac{\omega}{c}\Delta n\simeq\frac{\omega}{c}\frac{XY\sqrt{Y^{2} \sin ^{4} \theta/4+ (1-X)^{2}\cos ^{2} \theta }}{1-Y^{2}-X+X Y^{2} \cos ^{2} \theta }.  \label{Delta k}
	\end{equation}
	The polarization parameters $\mathcal{R}$ can be reduced to
	\begin{equation}
		\mathcal{R}=\frac{Y \sin ^{2} \theta}{2(1-X) \cos \theta}.  \label{reduce R}
	\end{equation}
	For a given frequency,  we can estimate the cyclotron resonance magnetic field $ B_{\rm res} $
	\begin{equation}
		B_{\rm res}=\frac{2 \pi m_{\rm e} c \nu}{e} \simeq 357~ \mathrm{G}\left(\frac{ \nu}{1 \mathrm{GHz}}\right),
	\end{equation}
	and the plasma resonance number density $ n_{\rm res} $ in the plasma medium
	\begin{equation}
		n_{\rm res}=\frac{ \pi m_{\rm e} \nu^2}{e^2} \simeq 1.2\times 10^{10} \mathrm{cm^{-3}}\left(\frac{ \nu}{1 \mathrm{GHz}}\right)^2.
	\end{equation}
	
	The astrophysical environment we discuss hereafter is supposed to be away from these resonance regions,  i.e., we adopt further approximations $ B\ll B_{\rm res} $ and   $ n_{e}\ll n_{\rm res} $. With these plasma parameters, following equations  (\ref{Delta k}) and  (\ref{reduce R}), the approximation of $ \Delta k $ can be treated as two cases  \citep{Melrose10}
	\begin{eqnarray}
		\Delta k\simeq\frac{\omega}{c}XY \left\{\begin{array}{ll}
			|\cos \theta|, \quad \quad 	 	|\cos \theta| \gtrsim Y/2 \quad \text{QL} ,
			\\
			\\
			Y\sin^2\theta / 2 , \quad 	|\cos \theta|\lesssim  Y/2 \quad \text{QT}.
		\end{array}\right.
	\end{eqnarray}
	
	Except for unusual circumstances, at high frequencies ($Y \ll 1$) the second case occupies only a very narrow range around  $\theta=\pi/2$, which we  refer to as the QT region,
	where $ \Delta k $ can be expected to be independent of $ \theta $. Within  the QT region, the natural mode is elliptically polarized, where it leads to the interconversion between CP and LP components, and we can define an FC rate to quantify this effect, which is
	\begin{eqnarray}
		\rho_{\rm L}=	\rho_{\rm Q}+ i 	\rho_{\rm U}&=&-	\Delta k\cos 2\chi_{\rm B}~e^{2i \psi_{\rm B}}\nonumber \\
		&\simeq &- \frac{\omega}{c}\frac{XY^2\sin ^{2} \theta}{2}e^{2i \psi_{\rm B} }.
	\end{eqnarray}
	
	On the other hand, in the QL region, the natural mode is almost circularly polarized and the modal axis is almost along the vertical axis of the Poincaré sphere. In this scenario, it only  changes  the polarization angle (PA) $ \psi $ of the incident EM wave, and similarly, we can define the Faraday rotation rate
	\begin{equation}
		\rho_{\rm V}=-	\Delta k\sin 2\chi_{\rm B}\simeq - \frac{\omega}{c}XY\cos\theta .
	\end{equation}
	Therefore, for a typical astrophysical environment, the  radiative transfer usually performs in the QL approximation, i.e. 	$  \rho_{\rm L}\ll \rho_{\rm V} $, with a difference of $Y/2$ between them.
	%--------------------------------------------------------
	\subsection{Toy model}
	Let us first consider a toy model in which the polarized radiation passes through the QL and QT regions  in turn, where the mode axis is along the V-axis in the QL region and along the Q-axis in the QT region.
	
	Given the initial polarization, the final polarization can be determined by
	\begin{eqnarray}
		&&	\left(\begin{array}{l}
			Q_{\rm f} \\
			U_{\rm f} \\
			V_{\rm f}
		\end{array}\right)=R(\theta_{\rm FC})R(\theta_{\rm FR})\left(\begin{array}{l}
			Q_{i} \\
			U_{i} \\
			V_{i}
		\end{array}\right), \label{transfer equation array}
	\end{eqnarray}
	with
	\begin{eqnarray}
		&&R(\theta_{\rm FC})R(\theta_{\rm FR})=\nonumber\\
		&&\left(\begin{array}{ccc}
			\cos \theta_{\rm FR}
			&~ -\sin \theta_{\rm FR}
			&~0
			\\
			\sin \theta_{\rm FR}\cos \theta_{\rm FC}
			&~		\cos \theta_{\rm FR}\cos \theta_{\rm FC}
			& ~-\sin \theta_{\rm FC}
			\\
			\sin \theta_{\rm FR}\sin \theta_{\rm FC}
			&~ 	\cos \theta_{\rm FR}\sin \theta_{\rm FC}
			&~ \cos \theta_{\rm FC}
		\end{array}\right),
	\end{eqnarray}
	where  the Faraday rotation  angle
	\begin{equation}
		\theta_{\rm FR}\equiv - \int d z	\rho_{\rm V}
		\equiv\left(\frac{\nu_{\rm FR}}{\nu}\right)^{2}
	\end{equation}
	is a physical parameter to characterize the Faraday rotation effect, this implies that a sufficiently significant Faraday rotation can occur  if the EM wave frequency is comparable to $ \nu_{\rm FR}  $ (i.e. $ \theta_{\rm FR}\sim 1~ {\rm rad} $), with
	\begin{eqnarray}
		\nu_{\rm FR}&\simeq& \left(\frac{\pi\nu_{\rm p}^{2} \nu_{\rm B}L\cos  \theta}{c}\right)^{1/2} \nonumber\\
		&=&27~\mathrm{GHz} \left(\frac{\Delta \mathrm{DM}}{0.1 \mathrm{pc}~ \mathrm{cm}^{-3}}\right)^{1/2}\left(\frac{B}{0.1 G}\right)^{1/2}.
	\end{eqnarray}
	Based on this equation, we can derive the expression for the rotation measure (RM)
	\begin{equation}
		\mathrm{RM} \equiv \frac{\theta_{\rm FR}}{2 \lambda^{2}}=8.1 \times 10^{5} ~\mathrm{rad}~ \mathrm{m}^{-2}\int \frac{d z}{\mathrm{pc}} \frac{n}{\mathrm{~cm}^{-3}} \frac{B_{z}}{\mathrm{G}}. \label{RM}
	\end{equation}
	Similarly, the FC angle is
	\begin{equation}
		\theta_{\rm FC}\equiv - \int d z \rho_{\rm L} \equiv	\left(\frac{\nu_{\rm FC}}{\nu}\right)^{3} 	
	\end{equation}
	is a  physical parameter that characterizes the Faraday  conversion  effect, and
	\begin{eqnarray}
		\nu_{\rm FC}&\simeq&\left(\frac{\pi\nu_{\rm p}^{2} \nu_{\rm B}^{2}L\sin ^{2} \theta}{2c } \right) ^{1/3}\nonumber\\
		&=&0.47 ~\mathrm{GHz} \left(\frac{\Delta \mathrm{DM}}{0.1 \mathrm{pc}~ \mathrm{cm}^{-3}}\right)^{1/3}\left(\frac{B}{0.1 G}\right)^{2/3}.
	\end{eqnarray}
	We note that $ \nu_{\rm FR}\gg \nu_{\rm FC} $, which suggests that Faraday rotation dominates in the usual magneto-ionic environment.
	Further, from equation (\ref{transfer equation array}), we can give an expression for the final CP
	\begin{equation}
		V_{\rm f}=	L_{\rm i}\sin \theta_{\rm FC}  \cos (\theta_{\rm FR}-	2\psi_{\rm 0})	+V_{\rm i}\cos \theta_{\rm FC}, \label{vf}
	\end{equation}
	where the LP component is $	L_{\rm i} =(Q_{\rm i}^2+U_{\rm i}^2)^{1/2} $. We can therefore conclude that the change in $ 	V_{\rm f} $ depends on whether the FC is significant as the  wave passes through the QT region.
	
	The PA is defined as $ 	\psi=1/2\arctan(U_{\rm f}/Q_{\rm f}) $.
	Substituting the specific expression for $ Q_{\rm f} $ and $ U_{\rm f} $, we notice that once the effect of the FC  becomes significant, the PA no longer has a simple power-law relationship with frequency, i.e. we cannot measure the RM in the traditional sense (\ref{RM}), this is because the presence of changes in $ V $ interferes the conversion between $ Q $ and $ U $. A common method for measuring RM is “QU fitting”, which models the Stokes Q and U quasi-periodic oscillations introduced by Faraday rotation \citep{Mckinven21, O'Sullivan12}.

	%--------------------------------------------------------
	\subsection{ The rQT region }
	For a more general discussion,  the magnetic field could change direction gradually, which suggests that the modal axes are changing orientation continuously.
	Let us  consider a region in whose center ($ z=0 $) the magnetic field along the line of sight suffers a sign reversal ($ \theta=\pi/2 $), which we call the rQT region, in which we can expand $ 	\rho_{\rm V} $ into $ 	\rho_{\rm V} =\rho_{\rm V}^{\prime}z$ and  $ 	\rho_{\rm L}$,	$\rho_{\rm V} $ can be assumed to be constant. We consider that the magnetic field and the ray are always lying in the same plane (non-twisted magnetic field).
	Without loss of generality, we can choose a plane such that $ \psi_{\rm B} = 0 $. Now  we define $f\equiv \rho_{\rm V}/	\rho_{\rm L}   $ and treat $ f=f(\tilde{z}) $ as a function of the dimensionless variable $ \tilde{z} = \rho_{\rm L} z $. The boundary of the rQT region $ [-\tilde{z}_0,\tilde{z}_0]  $ is determined by the condition  $f(\tilde{z})\gg 1$, which implies that the natural modes are nearly circular so that FC can be considered to be negligible at the boundary.

	Therefore, the representation of a polarized wave passing through an rQT region on a Poincaré sphere is that the modal axis gradually moves away from one pole of the sphere on one side  ($ z<0 $) of the rQT region, then crosses the equator of the sphere at the center  ($ z=0 $) of the rQT region and approaches the other pole of the sphere on the other side  ($ z>0 $) of the rQT region.
	The overall motion of the polarization state $\boldsymbol{ P}$ is  that it rotates at a constant latitude around this continuously evolving modal axis, which can only be determined by detailed calculations,  but it is also necessary to carry out a qualitative analysis before doing so.

	To proceed with the semi-quantitative calculation, we can define the modal coupling coefficient $C$  \citep{Melrose94,Melrose95}. Physically, $C$ is the ratio of the velocity ($ \left|\hat{\rho}^{\prime}\right| $) of the modal axis endpoint as it crosses the equator to the velocity ($ \left|\boldsymbol{\rho}\right| $) of the rotation of the polarization point around the modal axis.
	The case of $C\ll 1$ is denoted as weak coupling, in which the rotation of the polarization point is dominant  ($ \left|\hat{\rho}^{\prime}\right| \ll |\boldsymbol{\rho}|$). As  the modal axis turns from one pole of the sphere to the other, the polarization point rotates very fast so that the modal axis "drags" the polarization point and flips over together,
	which leads to the CP component $V$ changing its direction of rotation across the rQT region,  corresponding to the reversal of the $ V $ sign.
	The case $ C\gg 1 $ is denoted as strong coupling,  in which the modal axis rotates faster, the modes
	are tightly coupled at the center.  As the modal axis reverses its direction, the polarization point does not move significantly due to the relatively slow rotation of the polarization point around the modal axis. Therefore, the polarization point remains near its original position and the final result is that the CP changes by a small fraction. The case of intermediate coupling ($ C\sim 1 $), however, is very elusive, but it can be expected that a small change from the plasma parameters will lead to a significant change in the final polarization.

	Mathematically, a simple approximation of $C$ in the present context takes the form  \citep{Cohen60, Melrose10}
	\begin{eqnarray}
		C=\frac{\mathrm{d} f}{ \mathrm{d} \tilde{z}} \simeq \frac{\rho_{\rm V}^{\prime}}{\rho_{\rm L}^{2}}=\frac{4c}{\omega}\frac{1}{XY^3 \rm L_{\rm \theta}} \equiv\left(\frac{\nu}{\nu_{\rm T}}\right)^{4}. \label{couple}
	\end{eqnarray}
	The evolution of the polarization of a polarized wave after crossing such an rQT region depends entirely on the value of $ C $ at the center of the region,
	and it defines the transition frequency as
	\begin{eqnarray}
		\nu_{\rm T}	&=& \left(\frac{\pi}{2} \frac{\nu_{\rm p}^{2} \nu_{\rm B}^{3} L_{\theta}}{c}\right)^{1 / 4}
		\simeq 2.8~{\rm GHz}\left(\frac{B_{\rm 0}}{{100\rm G}}\right)^{3/4}\nonumber\\
		&\times& \left(\frac{n_{0}}{10^{4}~{\rm cm^{-3}}}\right)^{1/4}\left(\frac{{ L_{\theta}}}{R_{\odot}}\right)^{1/4} , \label{eq vT}
	\end{eqnarray}
	where $  n_0$ and $B_0 $ have the values taken from the center of the rQT region and $ \rm L_{\rm \theta}  $  is a characteristic length over  which the magnetic field significantly changes direction, i.e., the size of the rQT region.  For typical parameters, the transition frequency $ \nu_{\rm T} $ is expected to be in the radio band, 100 MHz-10 GHz, and a strong FC is expected at EM wave frequencies below $ \nu_{\rm T} $. The strong dependence of the coupling coefficient on frequency suggests that the transition from weak to strong coupling is very rapid as $\nu$ increases. It is noted that $ \nu_{\rm T} $ depends only weakly on the size of the rQT region and the plasma density in the rQT region, but has a strong dependence on the magnetic field strength.
	
	With the above definition and discussion, we can rewrite  Equation (\ref{transfer equation array0}) as
	\begin{equation}
		\frac{{\rm d} \boldsymbol{P}}{{\rm d} \tilde{z}}=(1,0, f(\tilde{z})) \times \boldsymbol{P},
	\end{equation}
	where
	\begin{equation}
		f(\tilde{z})=\frac{\rho_{\rm V}}{\rho_{\rm L}}=\frac{\rho_{\rm V}^{\prime}}{\rho_{\rm L}^2}\rho_{\rm L}z=C \tilde{z} .
	\end{equation}
	
\begin{figure*}
		\centering
		\subfigure[]{
			\includegraphics[width=0.4\linewidth]{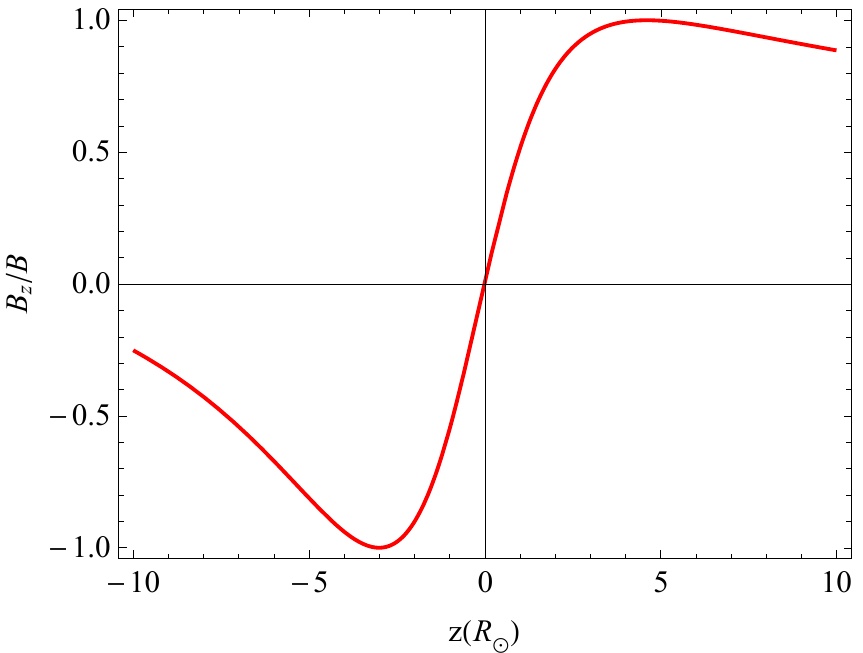}
			\label{fig:Bz}
		}
		\subfigure[]{
			\includegraphics[width=0.4\linewidth]{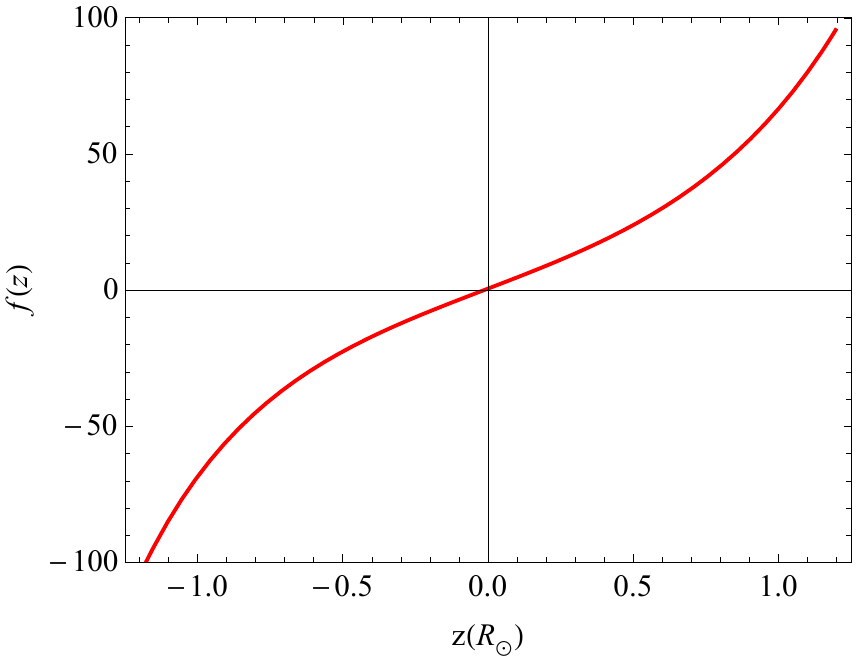}
		}
		\subfigure[]{
			\includegraphics[width=0.4\linewidth]{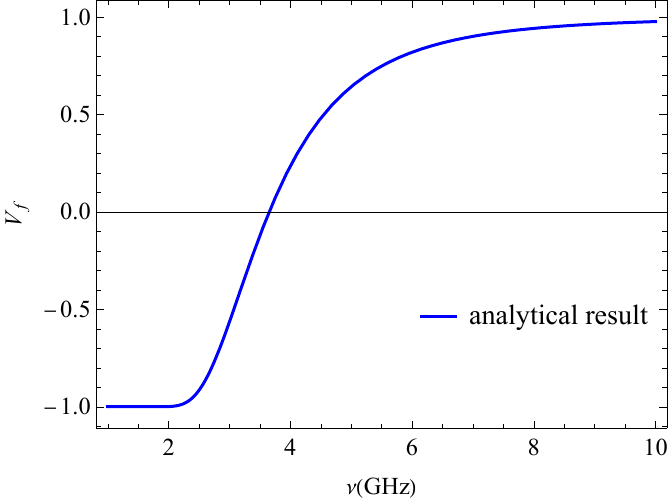}
			\label{fig analy}}
		\subfigure[]{
			\includegraphics[width=0.4\linewidth]{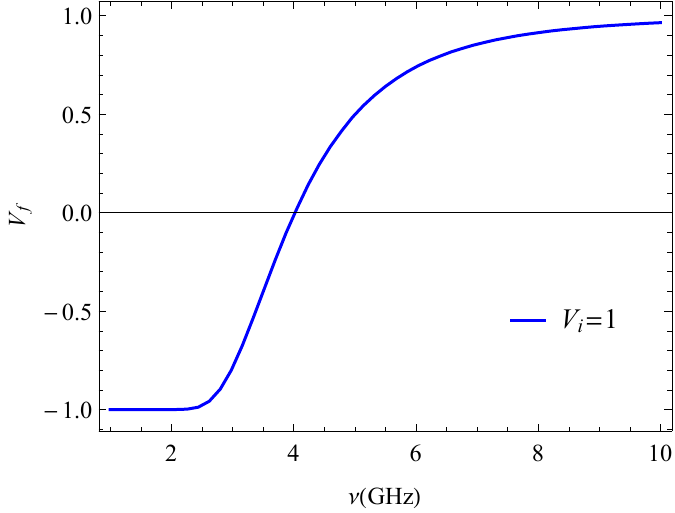}
			\label{fig:vi1}}
		\\
		\subfigure[]{
			\includegraphics[width=0.4\linewidth]{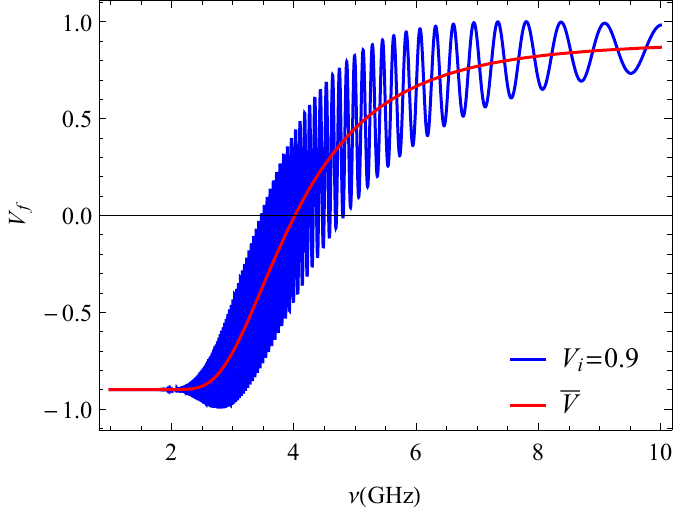}
			\label{fig:vi09}
		}
		\subfigure[]{
			\includegraphics[width=0.4\linewidth]{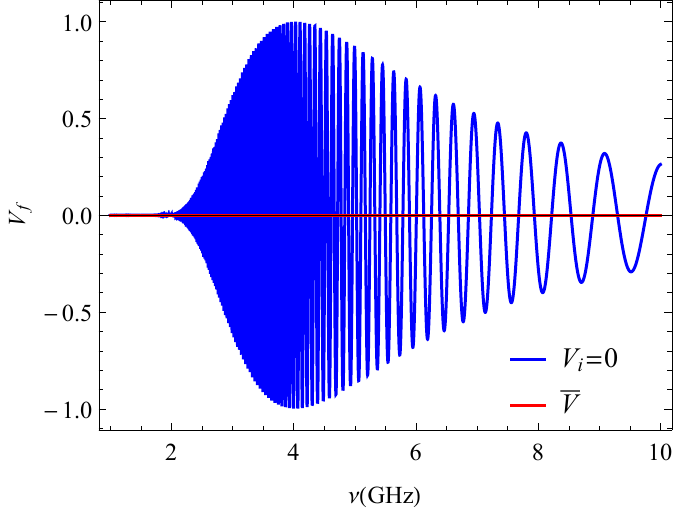}
			\label{fig:vi0}
		}
		\caption{(a) $ B_{z}/B $ as the function of $ z $ in the rQT region. $ z $ is on a scale of $ R_{\odot} $.
			(b) $ f(z) $ as the function of $ z $  near the rQT region center. It gives the region's  characteristic scale $ L_{\theta}\sim R_{\odot}$.
			(c) The analytical results obtained based on  Equation (\ref{Vf analy}). (d)
			$ V_{\rm f} $ as the function of  $\nu $ for $ V_{\rm i}=1$. (e)
			$ V_{\rm f} $ as the function of  $\nu $ for $ V_{\rm i}=0.9$. (f)
			$ V_{\rm f} $ as the function of  $\nu $ for $ V_{\rm i}=0$. Here the frequency range that we consider is 1-10 $ \unit{GHz}$, and the red solid line is the average value of the oscillations, which can be determined by Equation (\ref{Vf analy}).
			At the centre of the rQT region, there are $n_{0}=10^4 \unit{cm}^{-3}$, $B_{0}=100 \unit{G}  $, thus  $ \nu_{\rm T}\sim 3 {\rm GHz} $.
		 }
		\label{fig  numerical results}
	\end{figure*}

	The final average value\footnote{The average value refers to the average over a specific frequency range.} of the CP components $ V $ can be given analytically by its initial value
	\citep{Zheleznyakov64,Melrose94}
	\footnote{Their work considers only the case of 100$\% $ CP, and we have redefined their derived equations based on the numerical results in the  Fig. \ref{fig  numerical results}, as shown in  Equation (\ref{Vf analy}).}

	\begin{equation}
		\left\langle V_{\rm f}\right\rangle =\left(-1+2 \mathrm{e}^{-x}\right)  V_{\rm i}, \quad x=\frac{\pi}{2 C}. \label{Vf analy}
	\end{equation}
	
	Combining equation (\ref{couple}), by this analytical formula we draw a plot of $ V_{\rm f} $ versus frequency, as shown in Fig. \ref{fig analy}.
	Furthermore, for a strong coupling case  ($ \nu_{\rm T}\ll \nu $), following  Equation (\ref{vf}), one can perturbably determine the final CP as  \citep{Gruzinov19}
	\begin{eqnarray}
		V_{\rm f}&\simeq&  V_{\rm i}+ L_{\rm i} \int_{-{z}_0}^{{z}_0} d {z}	\rho_{\rm L} \cos  	\theta_{\rm FR} ({z})\nonumber\\
		&\simeq&  V_{\rm i}+ L_{\rm i}\times 2\sqrt{\frac{2\pi}{C}}\cos\left(\frac{\rho_{\rm V}^{\prime}}{2}z_0^2-\frac{\pi}{4}\right)F\left(z_0\sqrt{\frac{\rho_{\rm V}^{\prime}}{\pi}}\right)\nonumber \\
		&\simeq&  V_{\rm i}+ L_{\rm i}\times \left(\frac{\nu_{\rm CM}}{\nu}\right)^{2} \cos\left[ \left(\frac{\nu_{\rm RM}}{\nu}\right)^{2} -\frac{\pi}{4}\right].\label{Vf C>1}
	\end{eqnarray}
	where $ \theta_{\rm FR}({z})\simeq  \frac{\rho_{\rm V}^{\prime}}{2}({z}^2-{z}_0^2) $ is the Faraday rotation angle at an arbitrary location within the rQT region, and $F(x)$ is the Fresnel class C integral, denoting the real part of the Fresnel integral, with  $ F(\infty)=1/2 $.
	$ L_{\rm i} $, $ V_{\rm i} $ are the initial LP and CP, respectively.
	The above equation indicates that in the case of strong coupling, the final  CP oscillates with frequency and, surprisingly, its amplitude decreases with increasing initial CP.\footnote{In particular, if the incident radiation is completely circularly polarized, i.e. $  L_{\rm i}  $= 0, the oscillations disappear entirely, which reduces to the result in   \citep{Zheleznyakov64}.} This is confirmed by our numerical results in  Fig. \ref{fig  numerical results}.
	It is clear from the figure that $ V_{\rm f} $ reverses sign at the low-frequency end and rapidly recovers towards $ V_{\rm i} $ in the mid-frequency band, with a progressively smaller rate of oscillation, while at the high-frequency end, the oscillation behavior can be completely determined by Equation (\ref{Vf C>1}).
	
	$  \nu_{\rm CM}$ and $\nu_{\rm RM}$ determine the amplitude and frequency of the $ V_{\rm f} $ oscillation respectively, and their expressions are
	\begin{eqnarray}
		\nu_{\rm RM}&=&\left(\frac{\omega_{\rm p}^2\omega_{\rm B}{ L_{\theta}}}{8\pi^2c}\right)^{1/2} = 40.7 ~\unit{GHz}\left(\frac{B_{\rm 0}}{100\unit{G}}\right)^{1/2}\nonumber\\
		&\times& \left(\frac{n_{0}}{10^{4}~{\rm cm^{-3}}}\right)^{1/2} \left(\frac{{ L_{\theta}}}{R_{\odot}}\right)^{1/2},
	\end{eqnarray}
	
	\begin{eqnarray}
		\nu_{\rm CM}&=& \left(\frac{\omega_{\rm p}^2\omega_{\rm B}^3{ L_{\theta}}}{32\pi^3c}\right)^{1/4}= 4.5 ~\unit{GHz}\left(\frac{B_{\rm 0}}{100\unit{G}}\right)^{3/4}\nonumber\\
		&\times &\left(\frac{n_{0}}{10^{4}~{\rm cm^{-3}}}\right)^{1/4}\left(\frac{{ L_{\theta}}}{R_{\odot}}\right)^{1/4}=(2\pi)^{1/4}\nu_{\rm T},
	\end{eqnarray}
	\begin{eqnarray}
		\nu_{\rm RM}=\frac{\nu_{\rm  CM}^2}{\sqrt{\pi} \nu_{\rm B}}\simeq 201.6 ~\unit{GHz} \left(\frac{\nu_{\rm  CM}}{1\unit{GHz}}\right)^{2}\left(\frac{B_{\rm 0}}{\unit{G}}\right)^{-1}. \label{eq vRM}
	\end{eqnarray}
	
	This suggests that large amplitudes ($ 	\nu_{\rm CM} \simeq \nu $) of CP will typically induce very  drastic oscillations of CP ($ 	\nu_{\rm RM} \gg \nu $) for Gaussian-order magnetic field strength.
	Once the physical parameters at the center of the rQT region are given, we can adequately describe the polarization evolution of the EM wave.
	
	In summary, Faraday rotation is due to the magneto-radio wave propagation effect that is observed as a rotation of the plane of LP, where the rotation angle  $ \psi $ is linearly proportional to the square of the wavelength, with the slope being the familiar RM (Equation \ref{RM}). A common method of measuring RM is to fit the QU \citep{Mckinven21, O'Sullivan12}, which models Stokes quantities Q and U oscillations introduced by Faraday rotation,  and the measurement of CP is a direct calculation of $ V/I $. On the other hand, FC leads to interconversion between LP and CP, which is thought to be perhaps the origin of the CP observed by some radio sources. For FRBs, FC is thought to happen in special environments, such as (1) relativistic plasmas \citep{Vedantham19}; (2) pair plasmas \citep{Lyutikov22}; (3) there is a magnetic field reversal along the line of sight \citep{Gruzinov19, Qu23, li23}. FC has some unique observational features, such as the fact that CP and LP oscillate with frequency \citep{xu22}; CP profiles may be completely reversed after FC \citep{li23}.
	\citet{Lower21} and \citet{Kumar22} also provided a systematic study about how to measure FC.

	%--------------------------------------------------------
	\section{Binary environment}\label{sec:Binary}
	
	In this section, we will investigate a pulsar binary system with a companion star exhibiting a large-scale magnetic field. We find that the rQT region is readily encountered in binary systems.
	%--------------------------------------------------------	
	\subsection{Properties of companion stars}
	We define the mass of the companion star as $ M_{\rm c} $, the radius as $ R_{\rm c} $, and the mass loss rate as $ \dot{M} $.
	We consider the companion star's stellar wind as a radial, constant velocity outflow and attribute it to be the dominant contribution to the companion star's mass loss \citep{Mckee07}, then the electron number  density of the stellar wind at distance $ r $ from the star is given by
	\begin{equation}
		n_{\mathrm{w}}(r)=\frac{\dot{M}}{4 \pi \mu_{\rm m} m_{\rm p} v_{\rm w} r^{2}}=n_{\mathrm{w}, 0}\left(\frac{r}{R_{\rm c}}\right)^{-2},
	\end{equation}
	where $ \mu_{\rm m}  $ is the mean molecular weight of the stellar wind material, which can be assumed to be 1.29 for a typical stellar outflow with a hydrogen abundance of 0.7, and the mass loss rate depends on the type of companion star, ranges from $ 10^{-14} $ to  $ 10^{-5} M_{\odot} ~{\rm yr}^{-1}$. The $ n_{\mathrm{w}, 0}  $ is the electron number density of the star surface
	\begin{eqnarray}
		n_{\mathrm{w}, 0}& \simeq &8.2\times 10^{10} {\unit{cm^{-3}}}\left(\frac{R_{\rm c}}{1 R_{\odot}}\right)^{-3/2}\left(\frac{M_{\rm c}}{1 M_{\odot} }\right)^{-1/2}\nonumber\\
		&	\times &\left(\frac{\dot{M}}{10^{-8} M_{\odot} \mathrm{yr}^{-1}}\right).~
	\end{eqnarray}
	where the wind speed can be estimated as the escape velocity, i.e. $ v_{\rm w}=(2GM_{\rm c}/R_{\rm c})^{1/2} $.
	The stellar wind causes a modification in the magnetic field structure of the companion star. If the stellar wind is strong enough, it will straighten the magnetic field line, which will lead to the original dipolar magnetic field becoming radial. We can therefore define the Alfvén radius, which corresponds to the radius where the magnetic field pressure is equal to the ram pressure of the stellar wind   \citep{Yang22}, i.e. 	$ R_{\rm A}  \simeq (B_{\rm c}^{2} R_{\rm c}^{6}/{2\dot{M} v_{\rm w}})^{1 / 4}  $.
	Within the  Alfvén radius the magnetic field is dipolar, i.e. $ \boldsymbol{B}=B_{\rm c}R_{\rm c}^3(2\cos \theta \hat{\boldsymbol{r}}+\sin \theta \hat{\boldsymbol{\theta}}) /2r^3$, and outside this the field is radial, i.e. $  \boldsymbol{B}=B_{\rm c}R_{\rm c}^2 \hat{\boldsymbol{r}} /r^2$.  Here we ignore the toroidal field because the rQT region appears at the LOS only when the pulsar is at a specific position.
    Therefore, we consider that the magnetic field strength at a distance $ r $ from the highly magnetized companion star's center satisfies
	
	\begin{equation}\label{eq:B-field companion star}
		\begin{aligned}
			B&\simeq\left\{
			\begin{aligned}
				& B_{\rm c}\left(\frac{r}{R_{\rm c}}\right)^{-3}, \ &&r<R_{\rm A} , \\
				& B_c\left(\frac{R_{\rm A}}{R_{\rm c}}\right)^{-3}\left(\frac{r}{R_{\rm A}}\right)^{-2}, \ &&R_{\rm c}<R_{\rm A}<r.
			\end{aligned}
			\right.
		\end{aligned}
	\end{equation}

	\begin{figure*}
		\centering
		\subfigure[ LMCB.]{
			\includegraphics[width=0.4\linewidth]{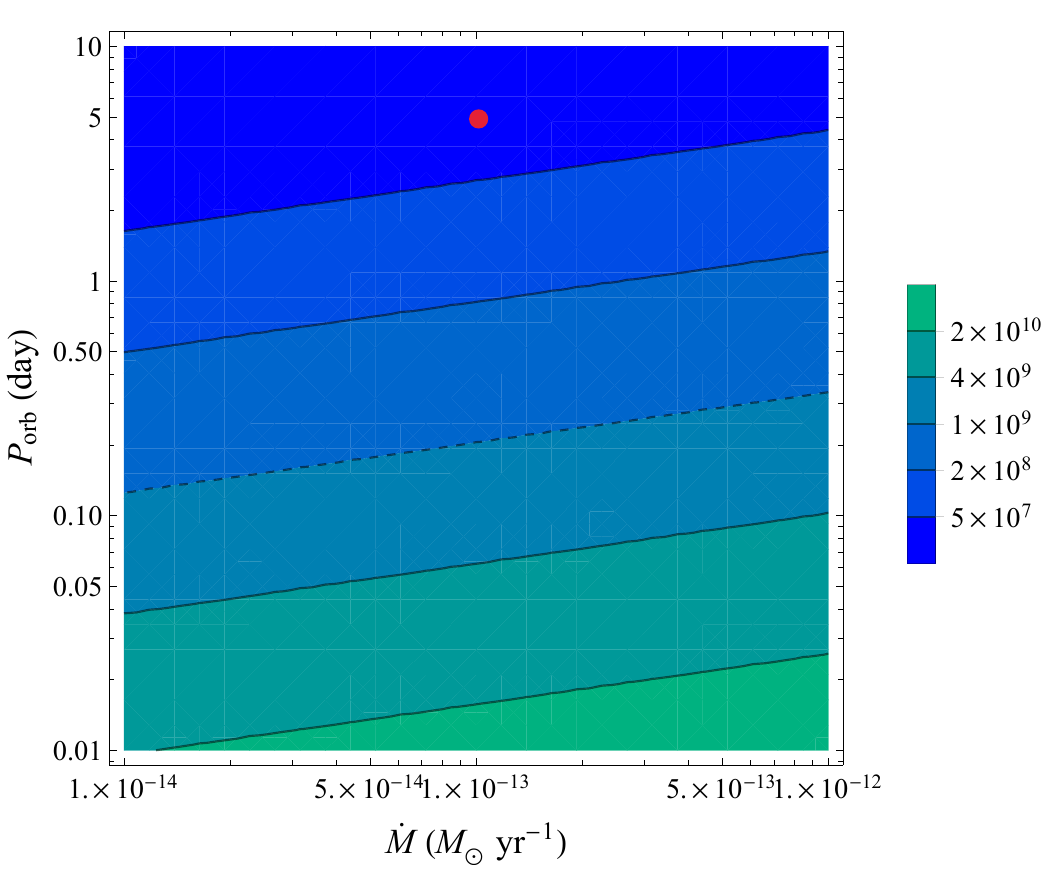}
			\label{fig LMXB}
		}
		\subfigure[ HMCB.]{
			\includegraphics[width=0.4\linewidth]{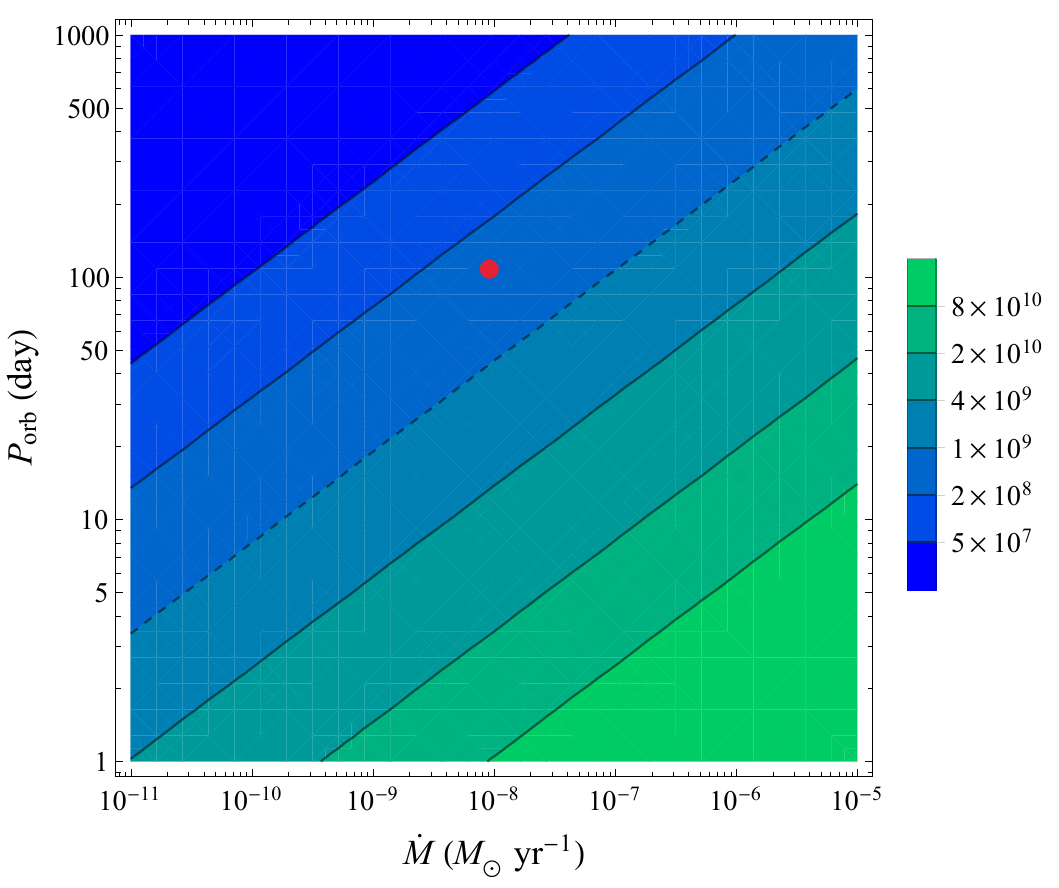}
			\label{fig HMXB}
		}
		\caption{The characteristic Faraday transition frequency  $\nu_{\rm T}$    as a function of the orbital period  $ P_{\rm orb}$   (day)  and the rate of mass loss $ \dot{M}(M_{\odot}\unit{yr}^{-1}) $ of the companion star. The dashed line corresponds to the radio frequency $ \nu \sim $ GHz, and strong FCs are expected in the region below this line. The red dots indicate the characteristic transition frequencies of the two binary systems.} \label{fig vt}
	\end{figure*}

	\begin{figure*}
		\centering
		\subfigure[$ P_{\rm c}=10 ~\rm day $.]{
			\includegraphics[width=0.4\linewidth]{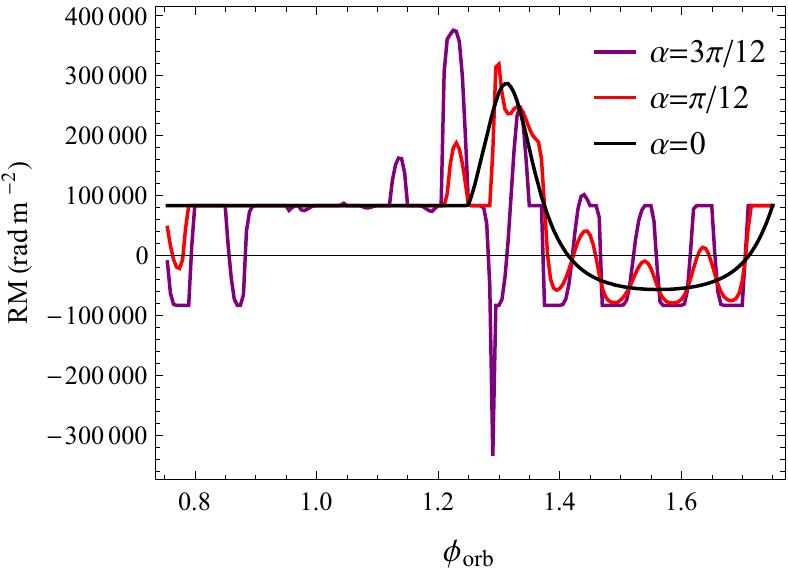}}
		\subfigure[$ P_{\rm c}=200 ~\rm day $. ]{
			\includegraphics[width=0.4\linewidth]{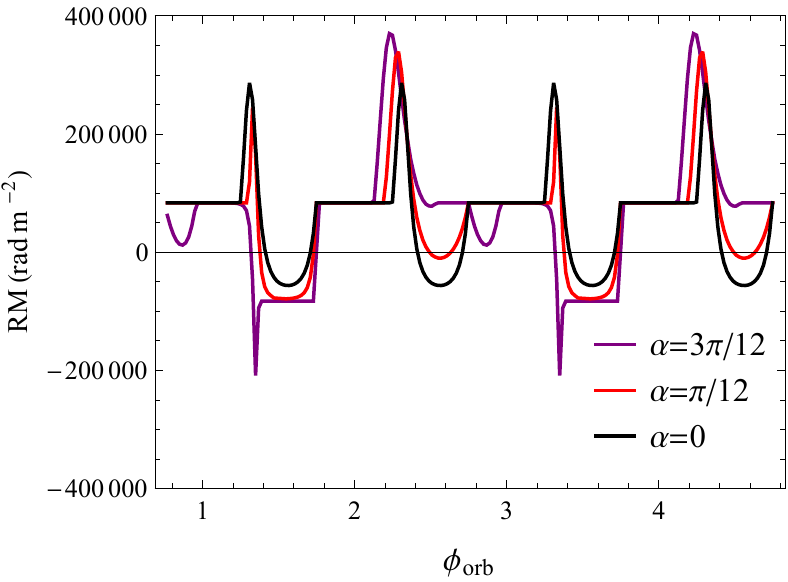}}
		\caption{The evolution of RM with orbital phase for different magnetic inclination angles $ \alpha $   in HMCBs. The pulsar is located  in superior conjunction at $\phi_{\rm orb}=1.25 $.}\label{fig RMHMXB}
	\end{figure*}

	\begin{figure*}
		\centering
		\subfigure[]{
			\includegraphics[width=0.4\linewidth]{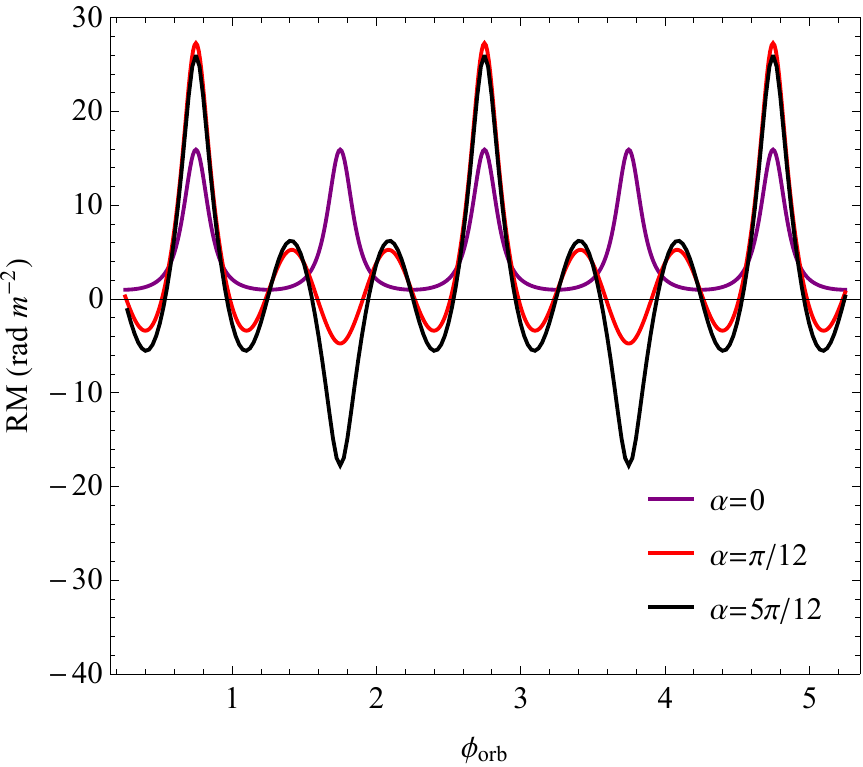}}
		\subfigure[]{
			\includegraphics[width=0.4\linewidth]{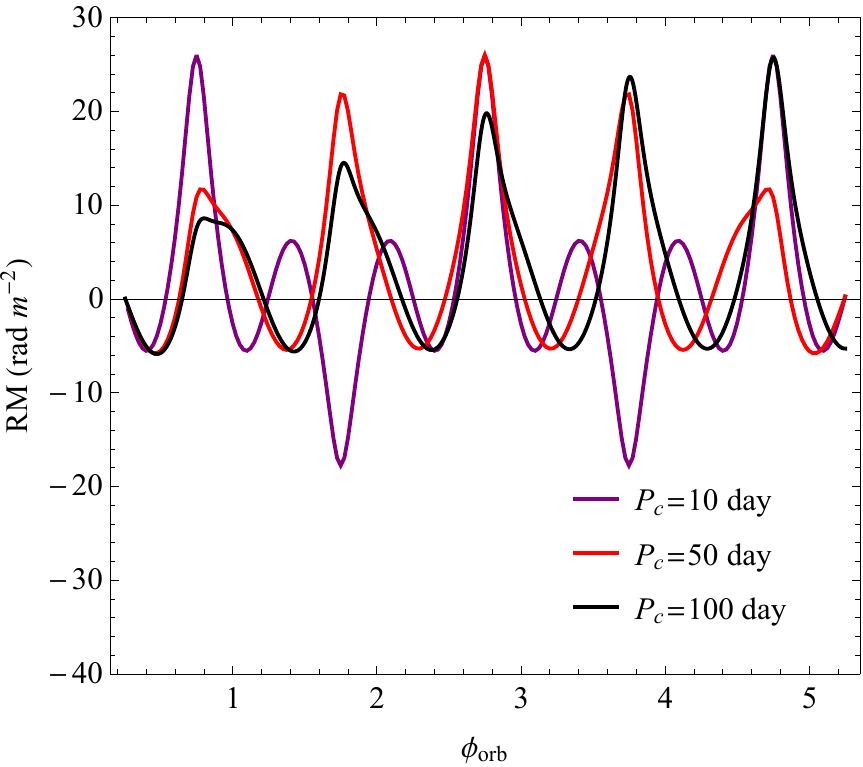}}
		\caption{The evolution of RM with orbital phase in LMCBs. (a) $P_{\rm c}=10 ~\rm day$, for different $\alpha=0, \pi/12 $ and $5\pi/12$.
			(b) $\alpha=5\pi/12$, for different $P_{\rm c}=$ 10, 50,  and 100 days over five orbital periods.}\label{fig RMLMXB}
	\end{figure*}
	
	\subsection{The characteristic transition frequency}
	In this work, we investigate the effect of companion stellar winds on the polarization properties of radio emissions in magnetized high-mass companion binary (HMCB) and low-mass companion binary (LMCB) systems involving an FRB source, respectively, whose parameter ranges are shown in Table \ref{t:range}. Before proceeding to the detailed calculations, it is useful to discuss qualitatively which type of binary systems are expected to cause a significant polarization evolution.

	\begin{deluxetable*}{lccccccccc}
		\tablenum{1}
		\tablecaption{Parameter ranges for the various properties of binary  systems.}
		\tablewidth{0pt}
		\tablehead{
			\colhead{Type\tablenotemark{a}} & \colhead{$M_{\rm{c}}$\tablenotemark{b}} & \colhead{$R_{\rm{c}}$ \tablenotemark{c}} & \colhead{$B_{\rm{c}}$\tablenotemark{d}} &
			\colhead{$\dot{M}_{\rm{c}}$\tablenotemark{e}} & \colhead{$P_{\rm{c}}$\tablenotemark{f}} & \colhead{$P_{\rm{orb}}$\tablenotemark{g}}  &  \colhead{References}\\
			\colhead{} & \colhead{($M_{\odot}$)} & \colhead{($R_{\odot}$)} & \colhead{(G)} &
			\colhead{($M_{\odot}\,\rm{yr^{-1}}$)} & \colhead{(day)} & \colhead{(day)}
		}
		%%\decimalcolnumbers
		\startdata
		HMCB
		& $10-100$ (16)          &
		$5-10  $  (6)     &
		$\sim 1-10^4$  ($ 10^3 $)  &
		$10^{-11}-10^{-5}$ ($ 10^{-8} $)  &
		$ 1 - 10^3$ (10)&
		$ 1 - 10^3$  (100)&
		1, 2, 3, 4, 5, 12\\
		LMCB
		& $ 0.1-7$ (0.5)      &
		$0.1-10 $  (0.5)       &
		$\sim 10^3$   ($ 10^3 $)     &
		$10^{-14}-10^{-12}$  ($ 10^{-13} $)  &
		$ 10 - 10^2$ (10)&
		$0.01 - 10$ (5)&
		6, 7, 8, 9, 10, 11\\
		\enddata
		\tablenotetext{\mathrm{a}}{\vspace{-4pt}Types of binaries.}
		\tablenotetext{\mathrm{b}}{\vspace{-4pt}The mass of the companion star.}
		\tablenotetext{\mathrm{c}}{\vspace{-4pt}The radius of the companion star.}
		\tablenotetext{\mathrm{d}}{\vspace{-4pt}The magnetic field of the companion star.}
		\tablenotetext{\mathrm{e}}{\vspace{-4pt}The mass loss rate of the companion star.}
		\tablenotetext{\mathrm{f}}{\vspace{-4pt}The spin period of the companion star.}
		\tablenotetext{\mathrm{g}}{\vspace{-4pt}The orbital period of the binary system.}
		\tablerefs{(1)  \cite{vac96}; (2) \cite{wad15}; (3) \cite{pul96}; (4) \cite{sno81}; (5) \cite{liu06}; (6) \cite{tak07}; (7) \cite{bor84}; (8) \cite{sur16};
			(9) \cite{liu07}; (10) \cite{pat21}; (11) \cite{Avakyan23}; (12) \cite{Fortin23}.
		}\label{t:range}
	\end{deluxetable*}
	
	According to Kepler's third law, we can derive the orbital radius
	\begin{eqnarray}
		R_{\rm orb} \simeq \left[ G\left(M_{\rm c}+M_{\mathrm{NS}}\right) P_{\mathrm{orb}}^{2} / 4 \pi^{2}\right]^{1 / 3}.
	\end{eqnarray}

	We refer to the catalog of Galactic X-ray binaries\footnote{Such binary systems are widely believed to contain neutron stars, but we don't require the main star to have X-ray emission, which is why we named it in a different way.} \citep{Fortin23, Avakyan23}, where the typical values of the binary parameters are those of maximum probability in the catalog.  $ M_{\rm c} $, $ R_{\rm c} $, and $ \dot{M} $ are dependent on the type of companion, and we first obtain typical values for $ M_{\rm c} $, and then use the empirical relationships to derive typical values for $ R_{\rm c} $ and  $ \dot{M} $ \citep{Eker18, Johnstone15}. The typical values\footnote{ With these typical binary parameters, the magnetized plasma satisfies $ B\ll B_{\rm res} $ and   $ n_{e}\ll n_{\rm res} $, exactly the weakly anisotropic medium we discussed in Sec \ref{sec:polarization}.} for the two types of binary parameters are listed in brackets of Table  \ref{t:range}. We note that there is a mass gap for the companion stars in the LMCB and HMCB, which is due to the fact that the parameter range was chosen with reference to the X-ray binary catalog.\footnote{ The lack of intermediate-mass companions in X-ray binary systems is mainly due to observational selection effect \citep{Chaty22}. These systems are usually weak X-ray emission sources. In contrast, high-mass X-ray binaries can be directly accreted by the stellar wind and low-mass X-ray binaries can undergo an effective accretion process through Roche-lobe overflow (RLO) to produce sufficiently strong X-ray emission.  In addition, when intermediate-mass
		X-ray binaries evolve towards RLO, this accretion phase only lasts for a short time and is thus not easy to detect.}

	For the LMCBs, we have $ R_{\rm A}\gg R_{\rm orb} $, so that the companion field is dominated by a dipole field, and there exists an rQT region near the magnetic axis, as shown in Fig. \ref{fig dipolar}. Then the characteristic transition frequency of the case can be estimated as
	
	\begin{eqnarray}
		\nu_{\rm T,LM}		
		& \simeq & 1.8 \times 10^{-2}~{\rm GHz}
		\left(\frac{B_{\rm c}}{1000 \mathrm{G}}\right)^{3/4}
		\left(\frac{R_{\rm c}}{0.5R_{\odot}}\right)^{13/8}
		\nonumber\\
		&\times&\left(\frac{M_{\rm c}}{0.5M_{\odot} }\right)^{-1/8}\left(\frac{\dot{M}}{10^{-13} M_{\odot}\mathrm{yr}^{-1}}\right)^{1/4}
		\left(\frac{R_{\rm orb}}{18R_{\odot}}\right)^{-7/4}.
	\end{eqnarray}
	
	For the HMCB, one has $ R_{\rm c}\ll R_{\rm A}\ll R_{\rm orb} $, so that the companion field is radially dominant, as shown in Fig. \ref{fig radial}, and  the characteristic transition frequency is
	
	\begin{eqnarray}
		\nu_{\rm T,HM}
		&\simeq& 0.5~{\rm GHz}
		\left(\frac{B_{\rm c}}{1000 \mathrm{G}}\right)^{3/8}
		\left(\frac{R_{\rm c}}{6R_{\odot}}\right)^{37/32}
		\left(\frac{M_{\rm c}}{16M_{\odot} }\right)^{-1/32}\nonumber\\
		&\times& \left(\frac{\dot{M}}{10^{-8} M_{\odot}\mathrm{yr}^{-1}}\right)^{7/16}
		\left(\frac{R_{\rm orb}}{205R_{\odot}}\right)^{-7/4}.
	\end{eqnarray}
	
	Here we assume that the size of the rQT region is proportional to the radius of its location ($ L_{\theta}\propto R_{\rm orb} $), due to the self-similarity of the field. In both cases above, the dependence of the characteristic transition frequency on the radius is very dramatic. %%The reason is that larger companion radii imply that the rQT region is further away from the companion and both the magnetic field and the electron number density decay very rapidly with increasing distance.
	For the adopted parameters, we can see that the LMCBs have a weak FC, while the magnetized HMCBs have a strong FC. The characteristic Faraday transition frequencies $ \nu_{\rm T} $ in the parameter space of orbital periods and mass loss rates are shown in Fig. \ref{fig vt}. The FC favors occurring under conditions of higher mass loss rates and shorter periods (this corresponds to a smaller orbital radius).

	In fact, strongly magnetized massive stars are usually rare. For example, less than 10$ \% $ of O stars are highly magnetized  \citep{wad15}, this may be due to the absence of convection zones around massive stars. Thus, massive stars lack magnetic dynamo mechanisms near their surfaces to convert convective and rotational energy into magnetic energy. Any strong magnetic field at its surface is either a remnant of past magnetic activity or originates from the interior of the star, and long-lived magnetic fields generated by these mechanisms appear to be rare.

	%--------------------------------------------------------	
	\subsection{Numerical calculations}\label{sec numerical}
	
	\begin{figure}
		\centering
		\includegraphics[width=\linewidth]{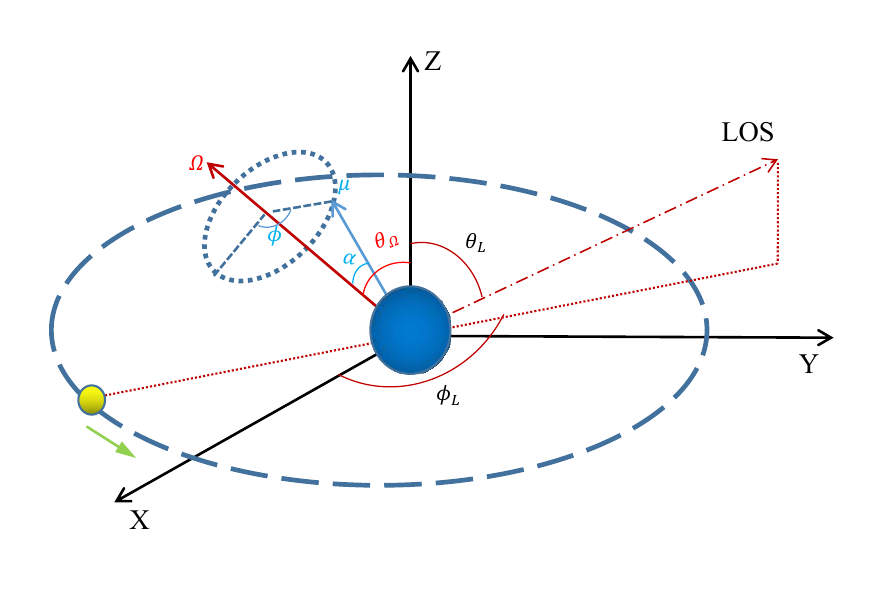}
		\caption{The geometric configuration of binary systems involving FRB sources, where the $XYZ$ coordinate system is chosen with the orbital plane lied in the $ XY $ plane and the rotation axis $\boldsymbol \Omega $ of the companion star lied in the $XZ$ plane. The angles with the $Z$ axis and the magnetic axes are denoted by $ \theta_{\Omega}  $ and $\alpha$ respectively.}	\label{fig binary}
	\end{figure}
	Next, we will solve the transfer equation (\ref{transfer equation array0}) numerically by substituting the specific environmental parameters of the binary star system\footnote{Some of the code for this section can be available on https://github.com/xiaygyg/LHMXB}.
	Let us now consider the binary system configuration illustrated in    Fig. \ref{fig binary}, where the $XYZ$ coordinate system is chosen so that the orbital plane  lies in the $ XY $ plane, the rotation axis $ \boldsymbol  \Omega $ of the companion star lies in the $XZ$ plane, and whose angles with the $Z$ axis  and the magnetic axes are denoted $ \theta_{\Omega}  $ and $\alpha$ respectively.
	The unit vector in the direction of the line of sight (LOS) can be denoted as

	\begin{equation}
		\hat{\boldsymbol l}=\left(\sin \theta_{\mathrm{L}} \cos \phi_{\mathrm{L}}, \sin \theta_{\mathrm{L}} \sin \phi_{\mathrm{L}}, \cos \theta_{\mathrm{L}}\right).
	\end{equation}
	
	For a typical LOS inclination of
	$\theta_{\rm L}=\pi/4$,
	we first plot the RM evolution with orbital phase $ \phi_{\rm orb}
	$\footnote{The  $ \phi_{\rm orb} $ is the normalisation of the orbital period, $\phi_{\rm orb} = 0.25$ for the pulsar is located in the superior conjunction.}
	for the magnetized HMCBs and LMCBs, as shown in Fig. \ref{fig RMHMXB} and \ref{fig RMLMXB}.
	For the magnetized HMCBs, we take the type parameters in  Table \ref{t:range}.
	We find that for strongly magnetized HMCBs, the companion stellar wind can be expected to contribute RM values of up to  $ 10^5 ~\mathrm{rad}~ \mathrm{m}^{-2}$ and that the RM undergoes sign reversal easily, as shown in Fig. \ref{fig RMHMXB}.
However, the extremely large RM values were not found in some of the known pulsars with high-mass companions \citep{Kaspi94,Stairs01,Lorimer06,Lyne15,Andersen23}. There are two reasons that high RM was not seen in these pulsars: (1) The high-mass companions associated with the six pulsars might have a lower surface magnetic field (e.g., $B \ll 1000~{\rm G}$), lower surface wind density ($n_{\rm w,0}\ll 10^8~{\rm cm^{-3}}$) or larger orbital period (e.g., $P_{\rm orb}\gg100~{\rm day}$), leading to a smaller RM value. (2) An extremely large RM can cause the depolarization of linearly polarized component due to the limited frequency resolution or the multipath propagation, which makes the extremely large RM unobservable.
	%which may be due to the Faraday depolarization effect in the companion wind. A large RM value means that the plane of linear polarization is rotating rapidly, and once the plane has rotated 360$^\circ$  within the telescopic observed resolution, depolarization occurs and RM is correspondingly undetectable.
	%This effect has been seen in the eclipsing Be-star system PSR B1259-63 \citep{Johnston96}. On the other hand, the companions we consider here are extremely magnetized, in fact, strong magnetic fields are rare for massive stars \citep{wad15}, and these pulsars do not show high RM values maybe because their companion's magnetic field is weak.

	On the other hand, for the left panel of Fig. \ref{fig RMHMXB},  the spin period of the companion star is shorter than the orbital period, and the evolution of RM is modulated mainly by the spin period; for the right panel, the orbital period is shorter, and the evolution of RM is modulated mainly by the orbital motion.  We also investigated the influence of the angle $ \theta_{\Omega} $ between the companion's rotation axis and the orbital plane, and found that the different $ \theta_{\Omega} $ simply introduced an asymmetry in the RM variations, which did not have much effect on the results. Further, one can expect to obtain very complex and diverse RM variations if we introduce eccentricity. Thus, we only consider the stellar wind contribution of the companion star, which is able to explain well the enriched and diverse RM variations (both in value and trend) in the recent FRB observations \citep{Hil21, mckinven22, xu22, Anna23},  and in particular to be advantaged in explaining significant RM reversal such as FRB 20190520B \citep{wang22}.
	
	%%However, since the rotation of the companion star only modifies the magnetic field configuration, this does not affect the  DM (which is only dependent on the electron number density and the distance of propagation). The DM  reaches its maximum value when the pulsar is at the superior conjunction; As the pulsar moves away from the superior conjunction, the DM decreases rapidly.
	For LMCBs, the RM contributed by the companion star winds is very small, but their evolution trends are also diverse, as shown in Fig. \ref{fig RMLMXB}.
	The substantial discrepancy in RM evolution between HMCB and LMCB is due to their different magnetic field geometry configurations.
	
	\begin{figure*}
		\centering
		\subfigure[]{
			\includegraphics[width=0.4\linewidth]{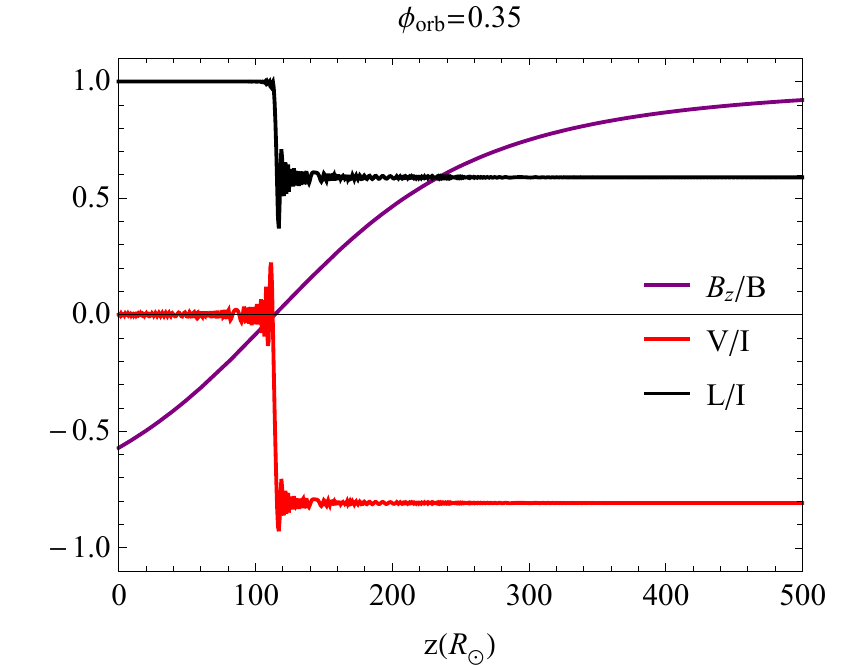}	}
		\subfigure[]{
			\includegraphics[width=0.4\linewidth]{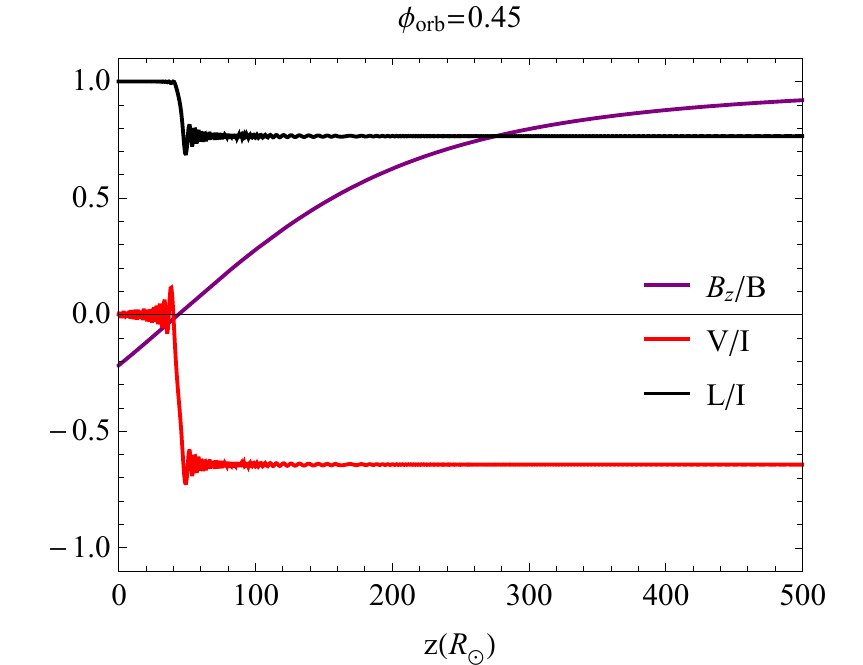}	}
		\caption{The evolution of polarization $(V,L)$ and $ B_{z} $  with distance along LOS at different orbital phases in HMCBs.
		} \label{fig VfHMXB1}
	\end{figure*}
	\begin{figure*}
		\centering
		\subfigure[]{
			\includegraphics[width=0.4\linewidth]{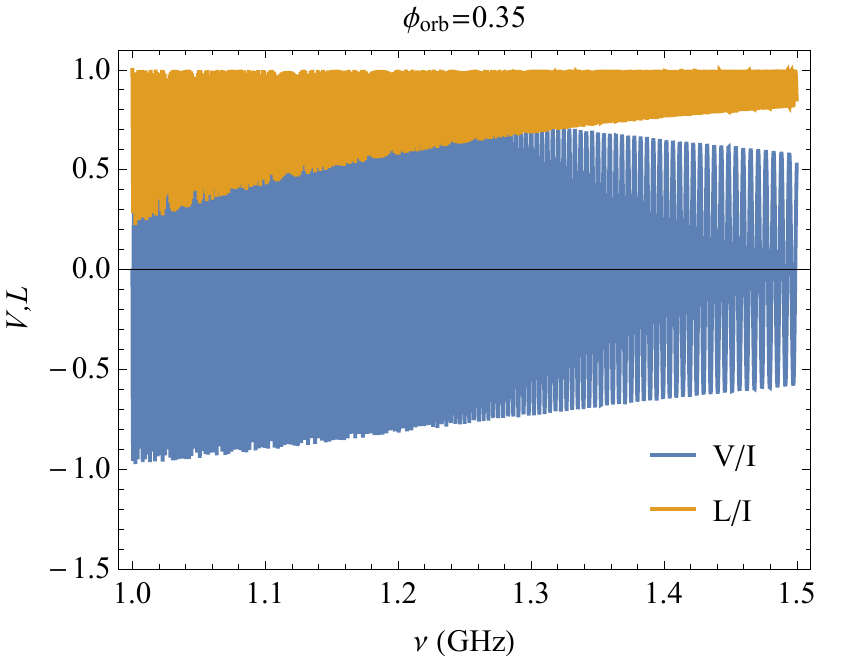}	}
		\subfigure[]{
			\includegraphics[width=0.4\linewidth]{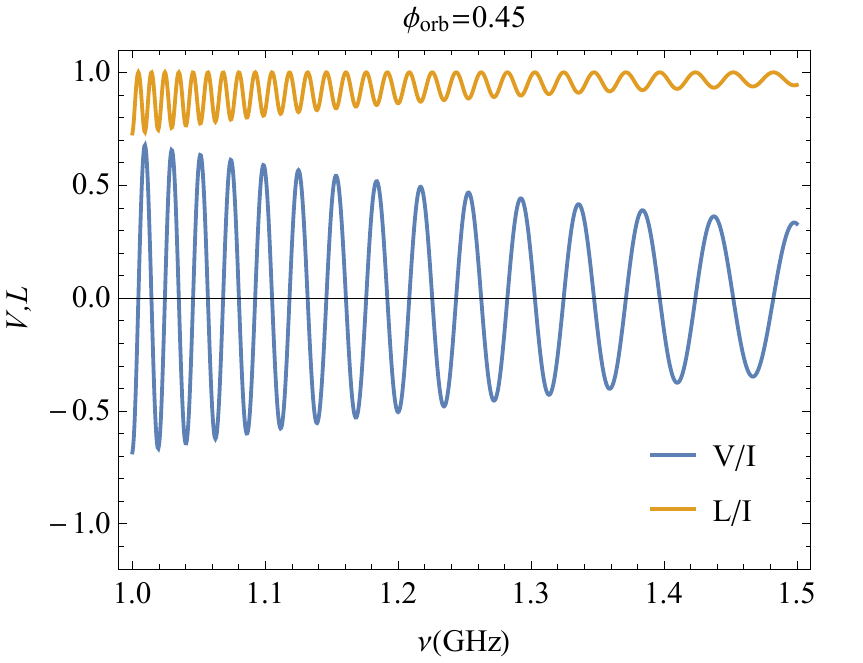}	}
		\caption{The evolution of polarization $(V,L)$ with frequency at different orbital phases in HMCBs.
		} \label{fig VfHMXB}
	\end{figure*}

	\begin{figure*}
		\centering
		\subfigure[]{
			\includegraphics[width=0.4\linewidth]{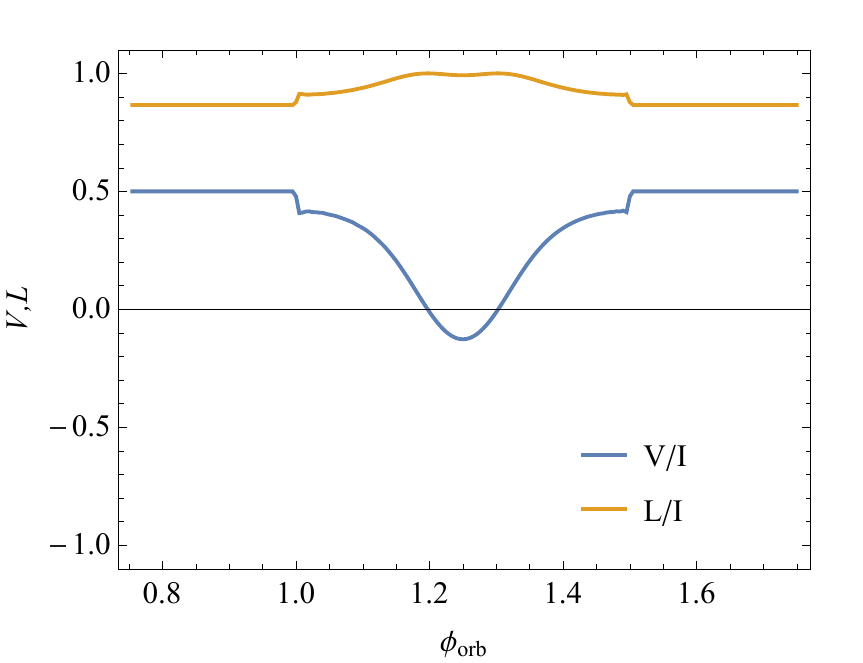}}
		\subfigure[]{
			\includegraphics[width=0.4\linewidth]{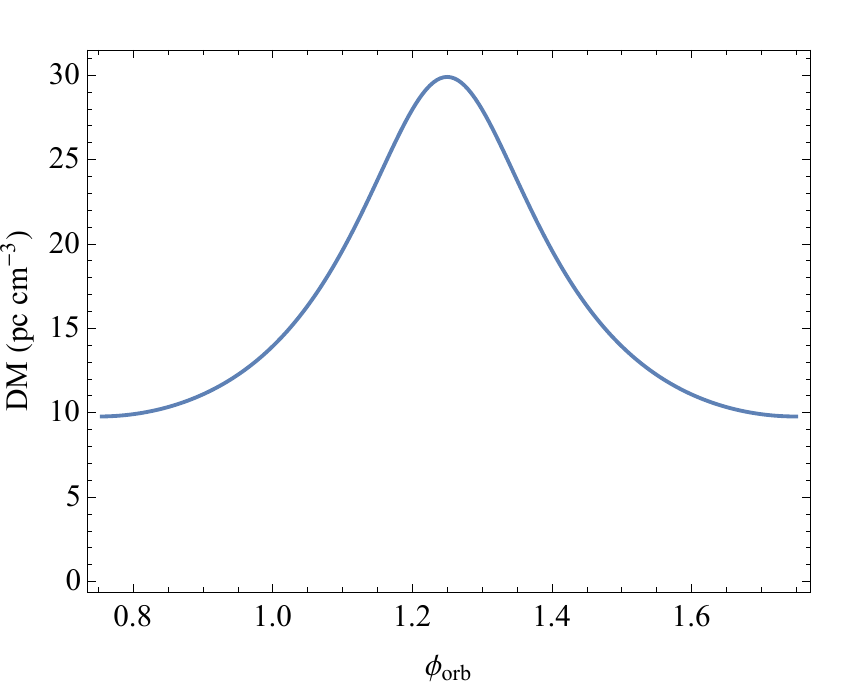}	}
		\caption{The evolution of polarization $(V,L)$ and DM with  orbital phases in HMCBs.
		} \label{fig VfHMXBphiorb}
	\end{figure*}
	
	Next, we investigate how the magnetic field of the companion star wind affects polarization. As shown in in Fig. \ref{fig VfHMXB1}, we note that after the polarized pulse passes through the companion's magnetic field, even if the pulse is completely linearly polarized at incidence, a CP component will be induced, which comes from the LP conversion, and this conversion only occurs with a magnetic field perpendicular to the LOS ($ B_{z}=0 $).
	As we consider the emitted pulse with a certain frequency bandwidth (1-1.5 {\rm GHz}, corresponding to the observed bandwidth of FAST), as shown in Fig. \ref{fig VfHMXB}, we find that the final CP  oscillates symmetrically   around the point with the CP degree equal to zero (initial CP) and that the oscillation decreases in rate and amplitude as the frequency increases. The oscillating behavior of the polarization with frequency alters significantly if the pulsar is in a different orbital phase. The FC becomes weaker farther away from the superior conjunction.

	In fact, for the HMCB, the CP oscillates with frequency at a very drastic rate due to its very strong plasma column density (the integration of the number density over the path of  LOS).
	If the oscillations are so fast that the period of oscillation is smaller than the telescopic observed resolution, this will probably lead to the  depolarization. Therefore, the FC in the HMCB is not favorable for producing significant CP. This may be the reason why CP in FRB 20190520B is very small and rare. On the other hand, according to Equation (\ref{eq vRM}), for the FC to induce a significant CP, a $\sim$100G perpendicular magnetic field at the QT region center is required. This will allow the CP to have a large amplitude ($\nu_{\rm  CM}$ $ \sim \rm GHz $) while its oscillation rate ($\nu_{\rm  RM}$ $\sim \rm GHz $) is not too fast.

	Further, we can also obtain the evolution of polarization with orbital phase, as shown in Fig. \ref{fig VfHMXBphiorb}.
	The polarization changes significantly only near the superior conjunction, the CP gradually becomes zero as the pulsar moves away from the superior conjunction, and then returns to its initial value at the normal phase.

\begin{figure*}
		\centering
		\subfigure[]{
			\includegraphics[width=0.4\linewidth]{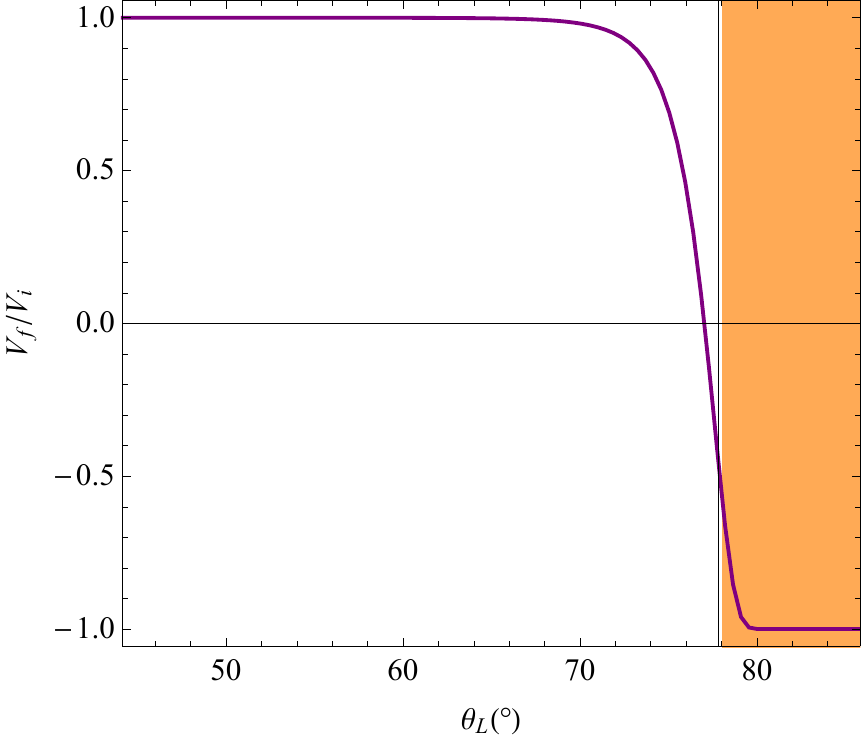}	}
		\subfigure[]{
			\includegraphics[width=0.4\linewidth]{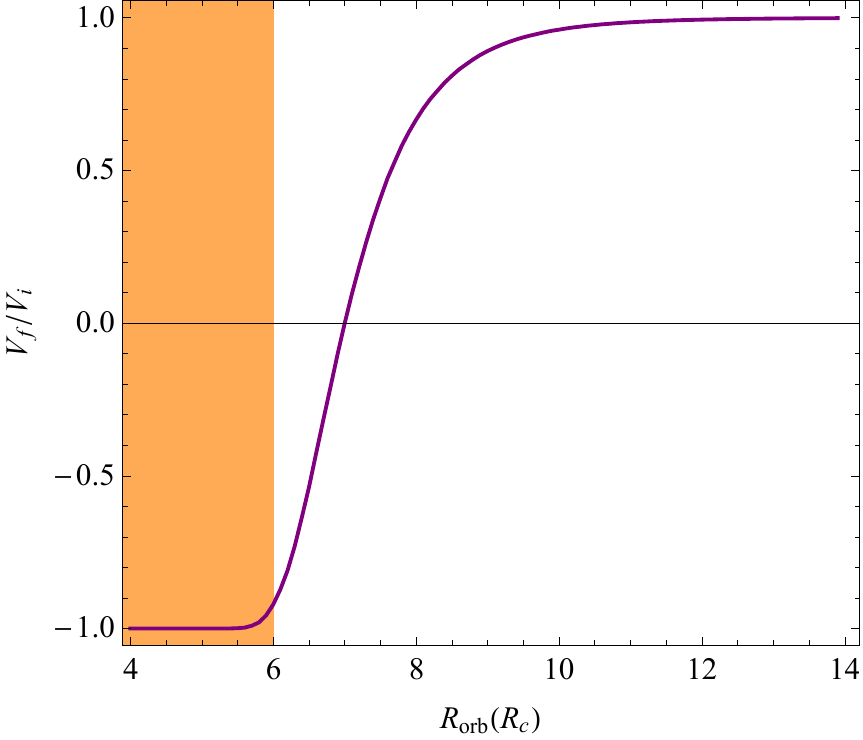}}
		\caption{(a) The value of $V_{\rm f}/V_{\rm i}$ as the function of  $ \theta_{\rm L} $ for $  R_{\rm orb} =16  R_{\rm c} $. (b) The value of $V_{\rm f}/V_{\rm i}$ as the function of  $ R_{\rm orb} $  for $ \theta_{\rm L} =\pi/4$.
			The strong FC can be generated even for LMCBs, as shown in the orange shaded region.}\label{fig LMXBvf}
	\end{figure*}

	Besides, we investigate the possibility that LMCBs can produce a strong FC by artificially altering the system's LOS $ \theta_{\rm L} $  and orbital radius  $ R_{\rm orb} $.  The requirements for producing a strong FC are shown in the orange-shaded region in Fig. \ref{fig LMXBvf}.
	Closer lines of sight to the orbital plane and smaller orbital radii can produce stronger FC, the binary system PSR B1744-24A \citep{li23} is one such example.

	\begin{figure*}
		\centering
		\subfigure[]{
			\includegraphics[width=0.4\linewidth]{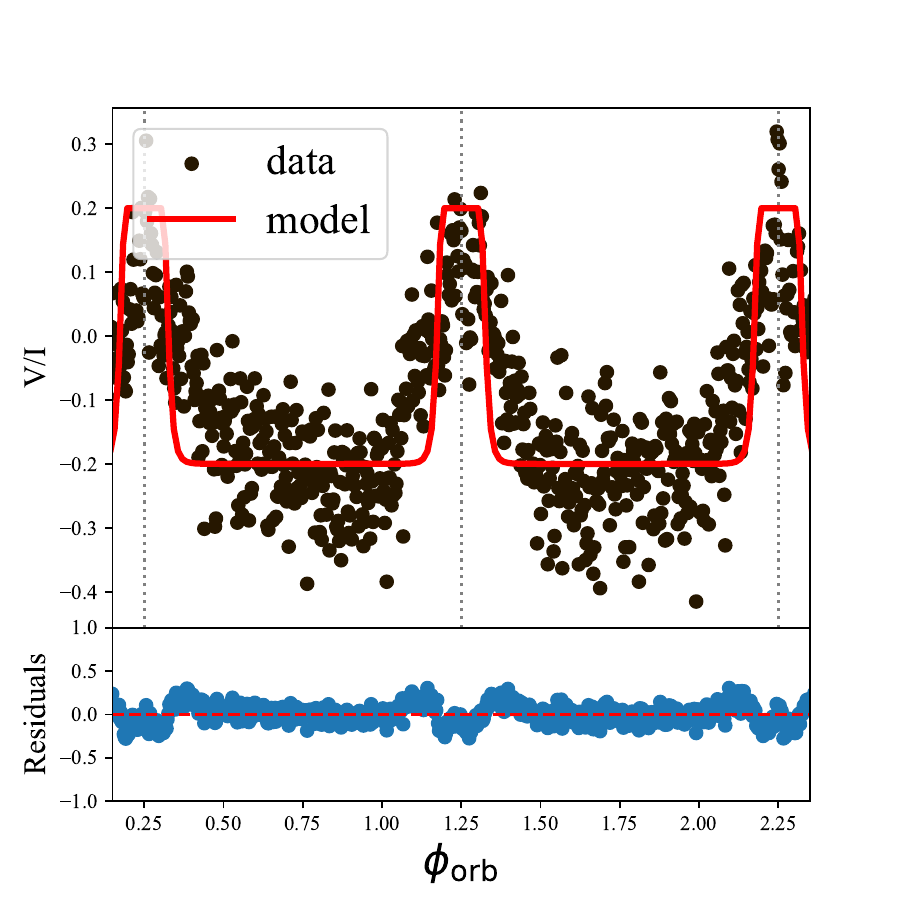}
		}
		\subfigure[]{
			\includegraphics[width=0.4\linewidth]{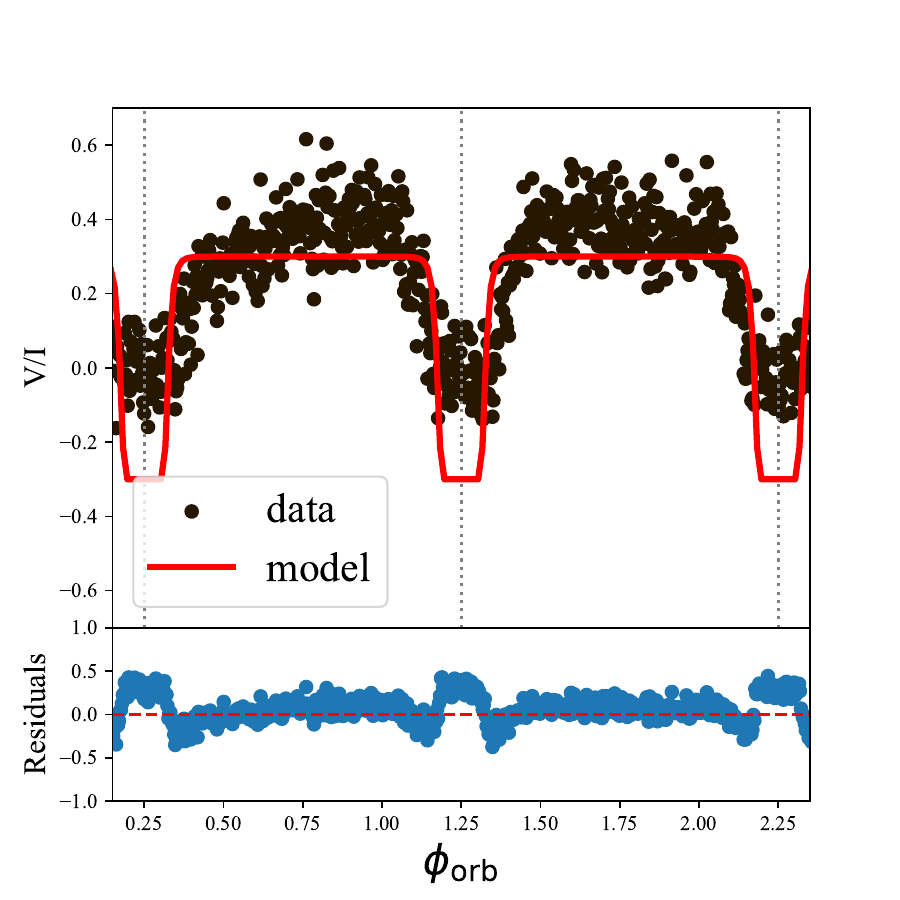}
		}
		\subfigure[]{
			\includegraphics[width=0.4\linewidth]{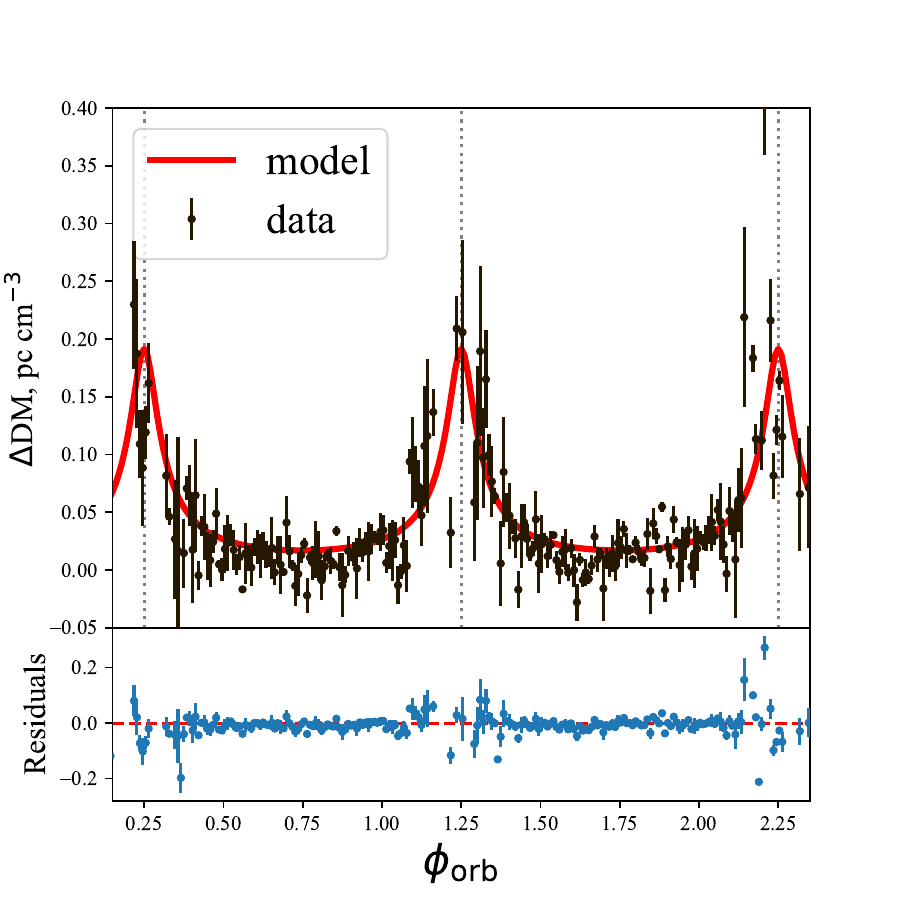}
		}
		\caption{Comparison between the model and observations of PSR B1744-24A.
			(a) The CP evolution with orbital phase near the spin phase 0.34.
			(b)  The CP evolution with orbital phase near the spin phase  0.32.
			(c) The DM evolution with orbital phase.
			The red line is the fitting line based on the model, and the black points are observation data.
			The residuals of the model and data are shown in the bottom panels, where the red horizontal dashed line is at zero residual.
			We take the binary parameters with   $ R_{\rm orb}=0.85R_{\odot} $, $ R_{\rm c}=0.12R_{\odot} $, $ B_{\rm c}=300 \unit{G} $, $ n_{\rm w,0}=4.5\times 10^{7}\unit{cm^{-3}} $, $ V_{\rm i}= 0.5 $, $ \theta_{\rm L}=5/12\pi $, $ \alpha =0 $.}
		\label{fig:model data}
	\end{figure*}
	
	\begin{figure*}
		\centering
		\subfigure[]{
			\includegraphics[width=0.4\linewidth]{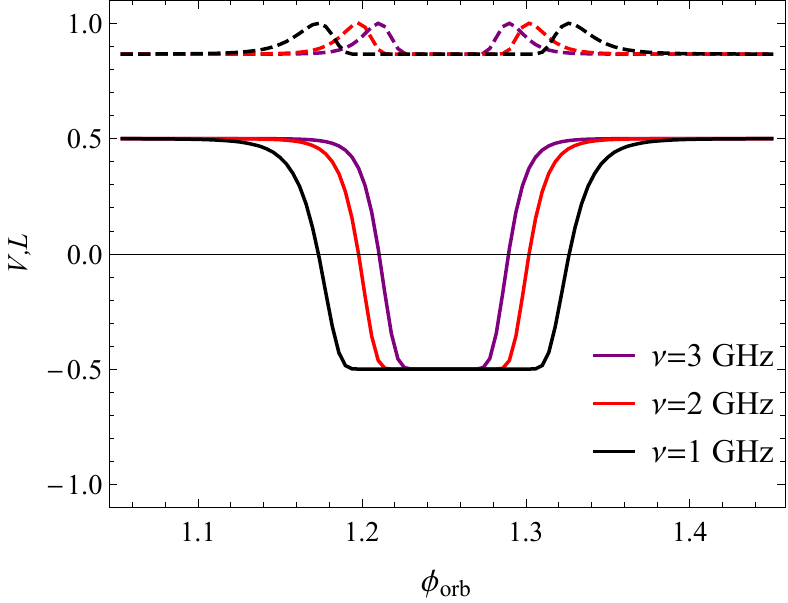}
			\label{fig Vphase}
		}
		\subfigure[]{
			\includegraphics[width=0.4\linewidth]{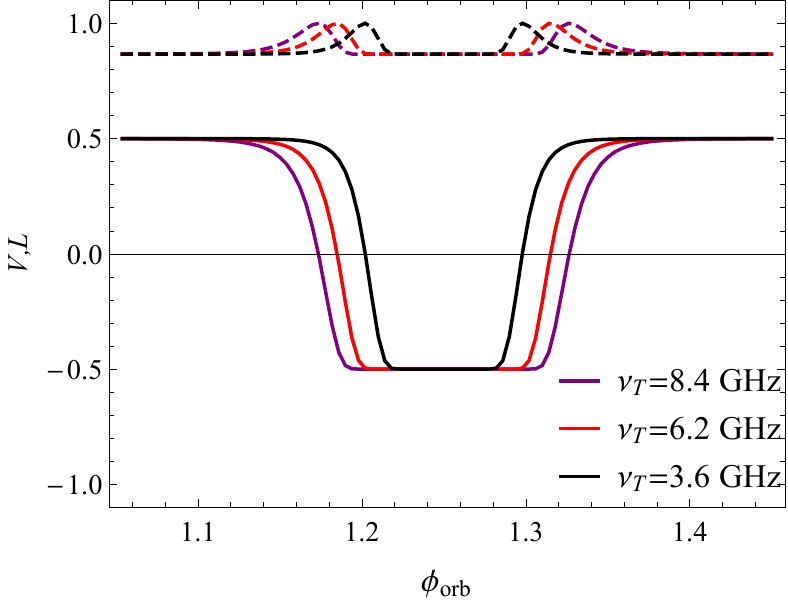}
		}
		\subfigure[]{
			\includegraphics[width=0.4\linewidth]{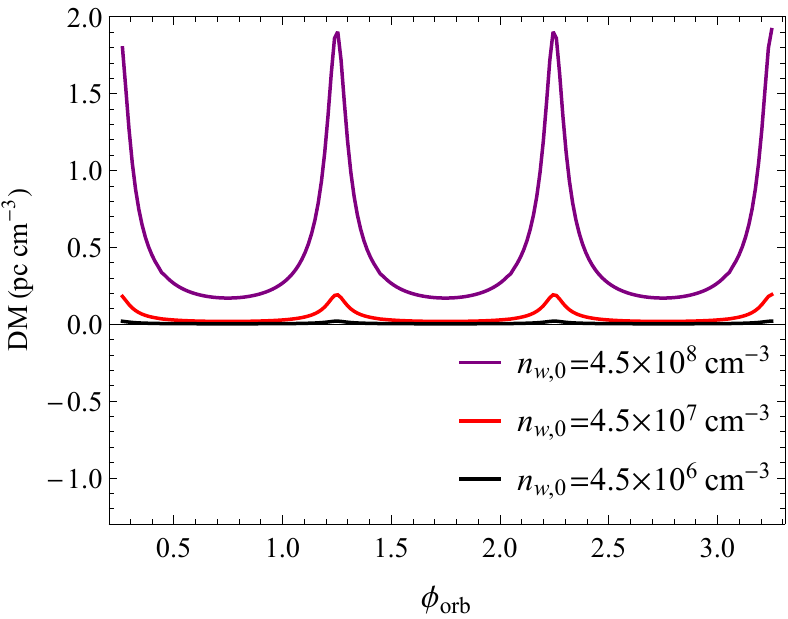}
		}
		\caption{
			The orbital evolution of linear/circular polarization degrees and DM. The panel (a) shows the evolution of linear/circular polarization degrees with different frequencies. The higher the frequency, the narrow the window for the Faraday conversion. The panel (b) shows the evolution of linear/circular polarization degrees with different $\nu_T$. The higher $\nu_T$, the wider the window.
			The panel (c) shows the DM evolution with different wind density at the companion surface. The larger $n_{\rm w,0}$, the larger the DM. }
		
		%(a), (b) The window for strong FC becomes narrower for higher observation frequencies $ \nu $ and transition frequencies $ \nu_{\rm T} $. (c) The smaller $ n_{\rm w,0}  $ leads to a smaller DM.
		
		\label{fig:sensitive}
	\end{figure*}

	\begin{figure*}
		\centering
		\subfigure[]{
			\includegraphics[width=0.4\linewidth]{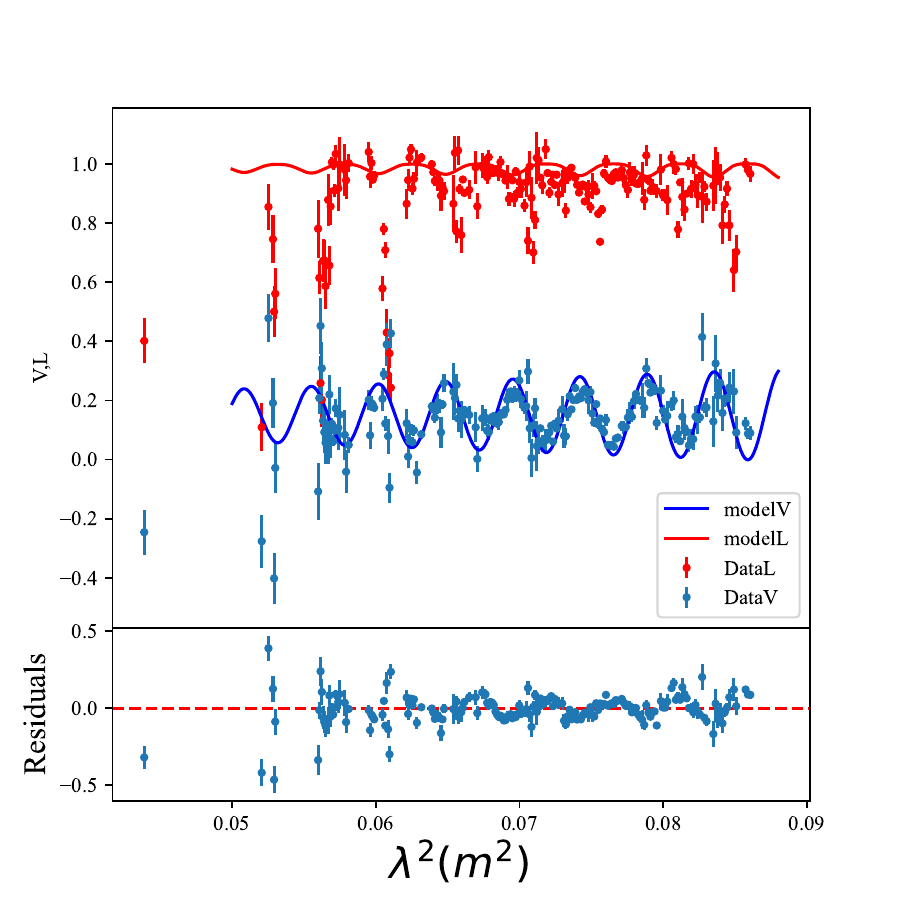}}
		\subfigure[]{
			\includegraphics[width=0.4\linewidth]{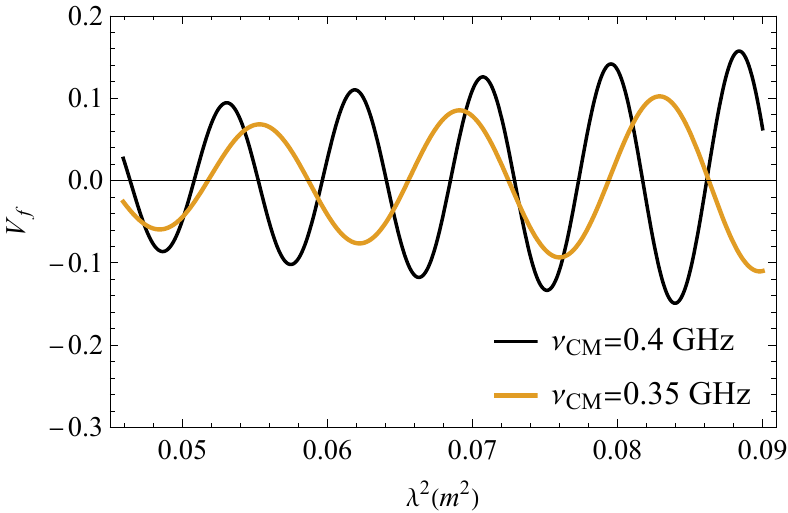}}
		\caption{
			Comparison between our model and observations of FRB 20201124A.
			(a) The observed oscillating polarization structure of the burst 926 of FRB 20201124A \citep{xu22}. The points represent the observation data, and the lines represent the fitting line with  $ \nu_{\rm  RM} \simeq 11.0~\rm GHz,~ \nu_{\rm  CM} \simeq 0.4~GHz$. Here the parameters are fiducial due to the model's complexity.
			(b) The outgoing CP as a function of  the squared wavelength based on Equation (\ref{Vf C>1}). Compared with the black line, the orange line has a smaller amplitude, leading to a smaller  oscillation rate.
		}\label{fig:vfxu}
	\end{figure*}

	%--------------------------------------------------------	
	\subsection{Discussion for the observations of PSR B1744-24A and FRB 20201124A}

	Observations of three consecutive orbital periods of the binary system PSR B1744-24A in the 1.5 GHz and 2 GHz bands show that the CP profile of the radio pulses near the superior conjunction  is completely opposite to that of the normal phase \citep{li23}, this provides strong evidence for FC. It emitted pulses that were also observed to  appear irregular large RM variation, which is very similar to some FRBs.
	These  observations suggest that the orbital motion of pulsar and the magnetic field of companion star play a decisive role. We applied our model\footnote{Here we have considered the magnetic axis perpendicular to the orbital plane, as the non-perpendicular case has only a minor effect on the results.} to  V and DM of PSR B1744-24A, as shown in Fig. \ref{fig:model data}, and give the residuals of the model with the data. The LP has a depolarization in many orbital phases, thus we neglected to overplot it.
	Notice that these results were obtained through numerical calculation via choosing appropriate fiducial parameters due to the model's complexity.
		% Notice that the displayed models are only for the fiducial parameters here, considering that our models are the result of numerical calculations.

	The reason that RM is not measured near the superior conjunction may be that the strong FC makes the variation of the Stoke parameters $Q, U$ with frequency no longer oscillating quasi-periodically with each other, but very chaotic, and therefore RM cannot be measured.

	Furthermore, we investigated how sensitive the model is to variations in the binary parameters in Fig. \ref{fig:sensitive}. The window of strong FC becomes narrower for higher observation frequencies $ \nu $ and lower transition frequencies $ \nu_{\rm T} $, owing to the coupling parameter $ C\propto (\nu/ \nu_{\rm T})^{4} $. The rQT region's plasma parameters $ n_{0} $ and $ B_{0} $ are coupled in $ \nu_{\rm T} $ (see Equation \ref{eq vT}). The DM drops sharply near the superior conjunction, due to the more dense plasma here. The smaller $ n_{\rm w,0}  $ leads to a smaller DM.
	
	However, it is noteworthy that there are other propagation effects that may arise as EM waves close to the companion, due to the large  magnetic field there, such as synchrotron-cyclotron absorption \citep{Qu23, li23}, which modulates the CP profile finely and also changes the total polarization fraction, which is not achieved by both Faraday rotation and FC.
	
	FRB 20201124A is the first repeating burst with significant CP. The CP is seen to oscillate periodically with frequency in some of these bursts  \citep{xu22}, which seems to be consistent with the strong coupling case discussed in our model. In particular, in this observation of FAST  \citep{xu22}, burst 779 has a smaller oscillating rate and amplitude compared to burst 926.
	This is consistent with our model that smaller oscillation amplitudes lead to smaller oscillation rates (Equation \ref{eq vRM}), as shown in Fig. \ref{fig:vfxu}.
	The large deviation of our model from LP data of burst 779 may be due to the presence of  additional absorption effects \citep{xu22, li23, Qu23} or depolarization.
	In particular, the fact that LP becomes less at lower frequencies is consistent with the picture of depolarization.

	%--------------------------------------------------------	
	\section{Summary and discussion} \label{sec:summary}
	In this paper, we investigate in detail the FC effect in cold plasmas and apply it to the binary system involving an FRB source. Then we analyze the polarization evolution of the radio bursts from the FRB source in the binary system.
	
	First,  we summarise the dispersion relation and polarization properties of the natural wave modes of polarized waves in different cold plasma environments (which can be described by $ \epsilon $)  and reveal the correlations between them (i.e.  Equations \ref{pm QL}, \ref{pm QT} and \ref{R/2}).
	By combining the transfer equation (\ref{transfer equation array0})  of the Stokes parameters, we depicted the polarization evolution due to the propagation of polarized waves in a plasma.
	Furthermore, we investigated the scenario for the idealized rQT region in the center of which the magnetic field  suffers a sign reversal.
	When the polarized waves pass through the rQT region, the natural wave modes will change from nearly circular to linear, and then from linear to the opposite circular.
	The FC effect can be generated when the EM wave passes through such a special region even in a clod plasma environment.
	Following the previous studies  \citep{Melrose95, Zheleznyakov64}, we found that the polarization of the outgoing radiation is only related to the environmental parameters at the center of the rQT region and to the size of the region, i.e.  Equation (\ref{eq vT}). We have studied this effect in detail and draw the following conclusions:
	\begin{itemize}
		\item
		For the weak coupling case ($ \nu_{\rm T}\gg \nu $), the  polarized waves crossing the rQT region lead to the CP component $V$ changing its direction of rotation, corresponding to the reversal of the $V$ sign.
		\item
		For the intermediate coupling case ($ \nu_{\rm T}\sim \nu $), the CP at outgoing would oscillate with frequency, and the average value of the oscillating CP degree $ \left\langle V_{\rm f}\right\rangle $ is smaller compared to the incident CP, $ V_{\rm i} $ because the CP component converts to an LP component  \citep{Zheleznyakov64}. It is worth noting that this effect does not proceed in reverse (Equation \ref{Vf analy}).
		\item
		For the strong coupling case ($ \nu_{\rm T}\ll \nu $, this is also the case satisfied by most binary systems), the CP  $V_{\rm f}$ is the addition of a term that oscillates periodically with frequency to the incident CP. The specific behavior of the oscillation is described by  Equation (\ref{Vf C>1}) and the amplitude of the oscillation is proportional to $L_{\rm i}$, which disappears as the incident CP is 100$\% $.
	\end{itemize}
	
	Further, based on the observations of PSR B1744-24A and FRBs, we consider a binary system with the companion star having large-scale magnetic fields, in which the rQT region is readily encountered as illustrated in Fig. \ref{fig rQT}.
	Further, we considered two kinds of binary systems, LMCB and HMCB. For their typical parameters, we found that the companion magnetic field of LMCB is usually dipolar, while that of HMCB is usually radial due to its strong  companion star wind. Next, we calculated their characteristic Faraday transition frequencies $ \nu_{\rm T} $ in both LMCB and HMCB, respectively, and we found that LMCB has a weak FC, while magnetized HMCB has a strong FC.
	The companions with higher mass loss rates and shorter periods favor the strong FC.
	
	For an strongly magnetized HMCB, when a fully linearly polarised FRB passes through the radial magnetic field of the companion star, its CP component will be induced and oscillates symmetrically around the point with the CP degree equal to zero, the rate and amplitude of the oscillation decrease as the frequency increases.
	The very strong plasma column density in the HMCBs can lead to CP oscillating with frequency at a very drastic rate, which may lead to depolarization.
	
	The oscillation behavior of the polarization with frequency alters significantly as the pulsar is in a different orbital phase.
	And the significant variation in polarisation only occurs near the superior conjunction, with the CP
	gradually tends towards zero and then returns to its value before incidence as the pulsar moves away from the superior conjunction.
	
	We also investigated the effect of the rotation of the companion star and found that a sufficiently significant RM reversal can be produced at large magnetic inclinations and that this variation in RM is very diverse. If the spin period of the companion star is shorter than the orbital period, the evolution of RM is modulated mainly by the spin period; if the orbital period is shorter, the evolution of RM is mainly modulated by the orbital motion.
	By introducing the eccentricity of the binary system and considering only the contribution of the companion wind, it is sufficient to fit well the complex and diverse RM variations observed in recent FRBs.
	However, since the rotation of the companion star only modifies the magnetic field configuration, this does not affect the DM. The DM value reaches the maximum when the pulsar is at the superior conjunction. As the pulsar moves away from the superior conjunction, the DM decreases rapidly.

	On the other hand, even typical LMCB also can produce strong FC if the LOS is sufficiently parallel to the orbit, the binary system PSR B1744-24A is one such example \citep{li23}. Observations of three consecutive orbital periods of the binary system PSR B1744-24A in the 1.5 GHz and 2 GHz bands show that the CP profile of the radio pulses near the superior conjunction is completely opposite to that of the normal phase \citep{li23}, this provides strong evidence for FC. We have thus obtained the evolution of polarization $(V,L)$ and DM with orbital phase (Fig. \ref{fig:model data}) for this binary, which can explain the observations of this binary system very well.
	Due to FC, the CP sign of the radio pulse near the superior conjunction is reversed.
	Furthermore, for higher frequencies, the window for the strong FC becomes narrower due to the coupling parameter $ C\propto \nu^{4} $.
	And we suggest that the reason that RM is not observed near the superior conjunction may be because the strong FC interferes with the periodic conversion between $Q$ and $U$, so that the PA no longer has a simple power-law relationship with frequency, i.e. we cannot measure RM in the conventional sense.
	
	In some  bursts of FRB 20201124A, the CP is seen to oscillate periodically with frequency \citep{xu22}, which is in good agreement with the strong coupling case discussed in our model. In particular, burst 779 has a smaller oscillating rate and amplitude compared to burst 926, which is consistent with our model's expectation, as shown in Fig. \ref{fig:vfxu}.

	In general, the polarization evolution due to the propagation effect in this model is independent of the FRB radiation energy, however, due to the very high luminosity of the FRBs, there might be non-linear effects near the FRB radiation region \citep{Lu20, yang20}, as well as the radiation pressure of the FRB may also interact with the environment medium \citep{yang21} and thus come to influence the polarization. There is a great deal of detail involved and it is beyond the scope of our work at present. A detailed analysis of these effects will be performed in the future.

	%--------------------------------------------------------	

	\acknowledgments
	
	We thank the anonymous referee for the detailed suggestions that have allowed us to improve this manuscript significantly. We also acknowledge helpful discussions with Kejia Lee, Dongzi Li, Ze-Nan Liu, Yong Shao, Yue Wu, Heng Xu, Zhen-Yu Yan, Bing Zhang, and Zhen-Yin Zhao. This work was supported by the National Key Research and Development Program of China (grant No. 2017YFA0402600), the National SKA Program of China (grant No. 2020SKA0120300), and the National Natural Science Foundation of China (grants 11833003, 12273009 and 11988101). Y.P.Y is supported by the National Natural Science Foundation of China grant No.12003028 and the National SKA Program of China (2022SKA0130100).

	\bibliographystyle{aasjournal}
	\bibliography{ref}

	\appendix
	
	\section{Cold plasma equations}\label{cold plasma equations}
	The propagation of EM waves in the cold plasma can essentially be thought of as an applied EM field that disturbs the particles at equilibrium in the plasma, and these disturbed particles can generate electric currents that in turn affect the EM field. The behavior of charged particles in the applied  EM field can be described by the particle motion equation and the continuity equation while  the EM field is governed by Maxwell's equations. They are
	
	\begin{eqnarray}
		\frac{\partial n_{\rm 1\alpha}}{\partial t}+\nabla \cdot\left(n_{\rm 0\alpha} \boldsymbol{u}_{\rm 1\alpha}\right) &=&0 ,\label{continuity}\\
		\frac{\partial \boldsymbol{u}_{\rm 1\alpha}}{\partial t} &=&\frac{q_\alpha}{m_\alpha}\left(\boldsymbol{E}_{1}+\frac{\boldsymbol{u}_{1\alpha} }{c}\times \boldsymbol{B}_{0}\right), \label{motion}\\
		\nabla \times\boldsymbol{E}_{1} &=&-\frac{1}{c}\frac{\partial \boldsymbol{B}_{1}}{\partial t}, \label{rot E}\\
		\nabla \times c \boldsymbol{B}_{1}- \frac{\partial \boldsymbol{E}_{1}}{\partial t} &=&4\pi\boldsymbol{j}=4\pi  \sum_\alpha  q_{\rm \alpha} n_{0\alpha} \boldsymbol{u}_{1\alpha}, \label{rot B}\\
		\nabla \cdot \boldsymbol{E}_{1} &=&4\pi \rho =4\pi \sum_\alpha q_{\rm \alpha} n_{1\alpha} ,\label{del E} \\
		\nabla \cdot \boldsymbol{B}_{1} &=&0 ,\label{del B}
	\end{eqnarray}
	with $ 	n =n_{0}+n_{1}, \quad \boldsymbol{B}  =\boldsymbol{B}_{0}+\boldsymbol{B}_{1}, \quad
	\boldsymbol{u}  =\boldsymbol{u}_{1}, \quad
	\boldsymbol{E}  =\boldsymbol{E}_{1}
	$	and $ n_{0}$, $ \boldsymbol{B}_{0} = B_0 \boldsymbol{e}_z  $ being constant in time and space. These  equations above are of nontrivial first-order form, which means we are only concerned with the small amplitude waves in this paper.
	
	We may write the dielectric tensor components  $ \varepsilon_{i j}=\delta_{i j}+\left(  4\pi i  /\omega\right) \sigma_{i j} $ by
	\be
	\varepsilon=\left(\begin{array}{ccc}
		S & -i D & 0 \\
		i D & S & 0 \\
		0 & 0 & P
	\end{array}\right), \label{varepsilon}
	\ee
	where the electrical conductivity $ \stackrel{\leftrightarrow}{\boldsymbol{\sigma}} $ is obtained by combining Equations (\ref{motion}) and (\ref{rot B}) with the microscopic Ohm's law $ \boldsymbol{j}= \stackrel{\leftrightarrow}{\boldsymbol{\sigma}}\cdot \boldsymbol{E} $, and the individual components in the dielectric tensor are defined as follows
	\begin{eqnarray}
		S&=&\frac{1}{2}(R+L),\label{S} \\
		D&=&\frac{1}{2}(R-L),\label{D}
		\\
		P &\equiv& 1-\sum_{\alpha} \frac{\omega_{\rm p \alpha}^{2}}{\omega^{2}} , \label{P}\\
		R &\equiv& 1-\sum_{\alpha} \frac{\omega_{\rm p \alpha}^{2}}{\omega\left(\omega+\omega_{\rm B\alpha}\right)}, \label{R}\\
		L &\equiv& 1-\sum_{\alpha} \frac{\omega_{\rm p \alpha}^{2}}{\omega\left(\omega-\omega_{\rm B\alpha}\right)},\label{L}
	\end{eqnarray}
	where the subscript $\alpha$ indicates the different components of the plasma and  $ \omega_{\rm B \alpha}  $ is the gyrofrequency for particles of type  $\alpha$, that is
	\begin{equation}
		\omega_{\rm B\alpha} \equiv \frac{q_{\alpha} B_{0}}{m_{\alpha} c}
	\end{equation}
	and
	\begin{equation}
		\omega_{\rm p\alpha} \equiv \frac{4\pi n_{0\alpha}q_{\alpha}^2 }{m_{\alpha} }
	\end{equation}
	is the plasma frequency.
	
	We need to introduce an additional plasma parameter $\epsilon$  \citep{melrose17}, that is defined as
	\begin{eqnarray}
		\epsilon \equiv \frac{	n_--n_+}{n_-+n_+},
	\end{eqnarray}
	which means that in the absence of positrons (for example, ion-electron plasma),  $\epsilon=1$ and in a charge-neutral pair plasma,  $\epsilon=0$.

	For a classical plasma composed of ions and electrons, at frequencies of EM waves much higher than the ion plasma $ \omega_{\rm pi}  $ and ion cyclotron frequencies $ 	\omega_{\rm Bi} $  (which is satisfied for most astrophysical environments),  the contribution of ions to the dispersion in the plasma is negligible compared to the contribution of electrons, because the mass of ions is much larger than the mass of electrons, and therefore ions are usually considered as stationary. Under the interaction of external EM fields, only the motion behavior of electrons is considered.
	In other words, the ion-electron plasma  with high-frequency EM wave propagation  can be treated as a cold pure electron gas, i.e., it can reduce to the case $\epsilon=1$.
	For simplicity, we then proceed to define two dimensionless parameters $X$ and $Y$ which incorporate $\omega_{\rm p}$ and $\omega_{\rm B}$
	\begin{equation}
		X=\omega_{\rm p}^2/\omega^2, \quad Y=\omega_{\rm B}/\omega, \label{eq:XY}
	\end{equation}
	where $\om_{\rm p}^2=4\pi e^2(n_++n_-)/m_{\rm e}$ is the total pair  plasma frequencies as well as $\omega_{\rm B}=eB/m_{\rm e}c $ is the value opposite to the negative electron cyclotron frequency.

	Following the definition of  Equations (\ref{R}) and (\ref{L}), we can write expressions for the fundamental components $ R $ and $ L $
	\begin{eqnarray}
		R=1-X(1+\epsilon Y)/(1-Y^2)=1-\frac{X}{1-Y^2} -\epsilon\frac{XY}{1-Y^2},\\
		L=1-X(1-\epsilon Y)/(1-Y^2)=1-\frac{X}{1-Y^2} +\epsilon\frac{XY}{1-Y^2}.
	\end{eqnarray}
	
	Further, we can give detailed expressions for the components P, S, and D of the dielectric tensor by the definition of  Equations (\ref{S}-\ref{P}), which respectively are
	\ba	S=1-\frac{X}{1-Y^2},
	\quad
	P=1-X,	\quad D=\frac{-\epsilon XY}{1-Y^2}.
	\label{cp1}
	\ea
	
	Finally, without loss of generality, we  may  choose a coordinate system  $ xyz  $ such that the refractive index vector $ \boldsymbol{n}=   (n \sin \theta, 0, n \cos \theta) $, so this leads to the wave equation  for the plane wave  $ \boldsymbol{E} \propto \mathrm{e}^{i(k \cdot r-\omega t)}  $, which has the expression
	\be	
	(\boldsymbol{n} \cdot \boldsymbol{E}) \boldsymbol{n}-n^{2} \boldsymbol{E}+\stackrel{\leftrightarrow}{\boldsymbol{\varepsilon}} \cdot \boldsymbol{E}=0,
	\ee
	where $ \boldsymbol{n}=c\boldsymbol{k} / \omega $.
	Then by substituting the dielectric tensor (\ref{varepsilon}) into the above equation, we are able that describe the wave equation in a matrix manner
	\be
	\left[\begin{array}{ccc}
		S-n^{2} \cos ^{2} \theta & -i D & n^{2} \cos \theta \sin \theta \\
		i D & S-n^{2} & 0 \\
		n^{2} \cos \theta \sin \theta & 0 & P-n^{2} \sin ^{2} \theta
	\end{array}\right]\left[\begin{array}{c}
		E_{x} \\
		E_{y} \\
		E_{z}
	\end{array}\right]=0 , \label{array E}
	\ee
	where $ \theta_{\rm B}  $  (for simplicity, the subscript B will be omitted throughout the following) is the angle between the magnetic field and the direction of EM wave propagation ($\boldsymbol{\hat{k}}$).
	By setting the determinant of the coefficients in the square matrix of  Equation (\ref{array E}) equal to zero, we can derive the dispersion relation for the cold plasma wave
	\be
	A n^{4}-B n^{2}+C=0 \label{n^4}
	\ee
	with
	\be
	\left\{\begin{array}{l}
		A=S \sin ^{2} \theta+P \cos ^{2} \theta ,\\
		B=R L \sin ^{2} \theta+P S\left(1+\cos ^{2} \theta\right), \\
		C=P R L.
	\end{array}\right.
	\label{ABC}
	\ee
	
	The general solution of the quadratic equation (\ref{n^4})  gives the dispersion equation
	\begin{equation}
		n^{2}=\frac{B\pm\left(B^{2}-4 A C\right)^{1 / 2}}{2 A }. \label{n^2}
	\end{equation}
	
	Physically, these two solutions correspond to the split of EM waves into two wave modes when they propagate in a plasma medium, which is referred to as the natural wave modes of the medium.
	In the magnetoionic theory (where the plasma is regarded as a cold magnetized pure electron gas), these two natural wave modes  are denoted as ordinary (O-mode) and extraordinary (X-mode) wave modes, and the detailed expressions for there are
	\begin{equation}
		n^{2}=\frac{2\left(1-Y^{2}\right)(1-X)\left(1-Y^{2}-X\right)-X Y^{2}\left(1-Y^{2}-X+\epsilon^{2} X\right) \sin ^{2} \theta\pm XY\sqrt{\Delta }}{2\left(1-Y^{2}-X+X Y^{2} \cos ^{2} \theta\right) \left(1-Y^2\right)} ,\label{pm n delta}
	\end{equation}
	with
	\begin{equation}
		\Delta=\left[\left(1-Y^{2}-X+\epsilon^{2} X\right)^{2} Y^{2} \sin ^{4} \theta+4 \epsilon^{2}(1-X)^{2}\left(1-Y^{2}\right)^{2} \cos ^{2} \theta\right]\label{Delta}.
	\end{equation}

	The two possibilities of the term $ \Delta  $ are
	\begin{eqnarray}
		\left(1-Y^{2}-X+\epsilon^{2} X\right)^{2} Y^{2} \sin ^{4} \theta \gg  4 \epsilon^{2}(1-X)^{2}\left(1-Y^{2}\right)^{2} \cos ^{2} \theta, \quad \text { QT, }  \label{pm QT}\\
		\left(1-Y^{2}-X+\epsilon^{2} X\right)^{2} Y^{2} \sin ^{4} \theta  \ll 4 \epsilon^{2}(1-X)^{2}\left(1-Y^{2}\right)^{2} \cos ^{2} \theta, \quad \text { QL. } \label{pm QL}
	\end{eqnarray}
	
	They correspond to the quasi-transverse  (QT) and quasi-longitudinal  (QL) approximations respectively  \citep{Stix92}.
	From the above we can see that for a pure pair plasmas (i.e. $ \epsilon=0 $), only the QT approximation can be satisfied, and in the section \ref{sec:polarization}, we can demonstrate that both  natural wave modes are completely linearly polarized with the QT approximation.

	\section{coordinate transformation} \label{sec coordinate trans}
	With the definition  in the section \ref{sec numerical}, we can give the expression for the magnetic moment $\boldsymbol{\mu}$ in the $XYZ$ system
	\begin{eqnarray}
		\boldsymbol{\mu}(\rm s)=\mu[(\cos \alpha \sin \theta_{\Omega}+\sin \alpha \cos \theta_{\Omega} \cos  \phi)\hat{\boldsymbol{X}}+(\sin \alpha \sin \phi)\hat{\boldsymbol{Y}}
		+(\cos \alpha \cos \theta_{\Omega}-\sin \alpha \sin \theta_{\Omega}\cos \phi)\hat{\boldsymbol{Z}}].
	\end{eqnarray}
	
	The polar angle $ \theta_{\mu} $ and the azimuthal angle $ \phi_{\mu} $ of $\boldsymbol{\mu} $ can therefore be derived
	\begin{eqnarray}
		\tan \phi_{\mu} &=&\frac{\sin\alpha \sin \phi}{\cos \alpha \sin \theta_{\Omega}+\sin \alpha \cos \theta_{\Omega} \cos  \phi}, \\
		\tan \theta_{\mu} &=&\frac{\sin\alpha \sin \phi}{(\cos \alpha \cos \theta_{\Omega}-\sin \alpha \sin \theta_{\Omega}\cos \phi)\sin \phi_{\mu}},
	\end{eqnarray}

	For a dipole field, in the coordinate system $ X^{\prime}Y^{\prime}Z^{\prime} $ where the magnetic moment $ \boldsymbol{\mu} $ along the $ Z^{\prime} $-axis, the magnetic field component is in the form
	\begin{eqnarray}
		B_{x}^{\prime} &=& \frac{3B_{\rm c}R_{\rm c}^3}{2 r^{3}}  \sin \theta \cos \theta \cos \varphi, \\
		B_{y}^{\prime} &=& \frac{3B_{\rm c}R_{\rm c}^3}{2 r^{3}}  \sin \theta \cos \theta \sin \varphi, \\
		B_{z}^{\prime} &=& \frac{B_{\rm c}R_{\rm c}^3}{2 r^{3}}\left(2 \cos ^{2} \theta-\sin ^{2} \theta\right).
	\end{eqnarray}
	
	For a radial field,
	\begin{eqnarray}
		B_{x}^{\prime} &= & \operatorname{sgn}(z)\frac{B_{\rm c}R_{\rm c}^2 }{ r^{2}}  \sin \theta\cos \varphi, \\
		B_{y}^{\prime} &= & \operatorname{sgn}(z)\frac{B_{\rm c}R_{\rm c}^2 }{ r^{2}}  \sin \theta \sin \varphi, \\
		B_{z}^{\prime} &= & \operatorname{sgn}(z)\frac{B_{\rm c}R_{\rm c}^2 }{ r^{2}} \cos \theta,
	\end{eqnarray}
	where $\operatorname{sgn}(z)$ is  the sign function.
	For a toroidal field
	\begin{eqnarray}
		B_{x}^{\prime} &= &-\frac{B_{\rm c}R_{\rm c} }{ r}  \sin \varphi ,\\
		B_{y}^{\prime} &= &\frac{B_{\rm c}R_{\rm c} }{ r}   \cos \varphi ,
	\end{eqnarray}
	where
	$ (r,\theta,\varphi ) $ gives the radial distance, polar angle, and azimuthal angle, which can be obtained from their Cartesian coordinates  according to
	\begin{eqnarray}
		r&=&\sqrt{x^{2}+y^{2}+z^{2}}, \\
		\theta&=&\arccos \frac{z}{\sqrt{x^{2}+y^{2}+z^{2}}},\\
		\varphi&=&\operatorname{sgn}(y) \arccos \frac{x}{\sqrt{x^{2}+y^{2}}}.
	\end{eqnarray}
	
	The coordinate transformation between the two  systems is given by
	\begin{eqnarray}
		\left(\begin{array}{c}
			X^{\prime} \\
			Y^{\prime} \\
			Z^{\prime}
		\end{array}\right)=\left(\begin{array}{ccc}
			\cos \theta_{\mu} \cos \phi_{\mu} & ~-\cos \theta_{\mu} \sin \phi_{\mu} &~ \sin \theta_{\mu} \\
			\sin \phi_{\mu} & ~\cos\phi_{\mu} & ~0 \\
			-\sin \theta_{\mu} \cos \phi_{\mu} &~ \sin \theta_{\mu} \sin \phi_{\mu} &~ \cos\theta_{\mu}
		\end{array}\right)\left(\begin{array}{l}
			X \\
			Y \\
			Z
		\end{array}\right)
		\label{coordinate transformation},
	\end{eqnarray}
	where the square matrix in  Equation (\ref{coordinate transformation}) can be denoted $M$, which means the rotation matrix of the coordinate transformation so that the magnetic field components in the $XYZ$  system can be derived by     $  {\boldsymbol{B} }=M^{-1} {\boldsymbol{B}^{\prime}} $.

\end{document}